\documentclass[twoside,11pt,final] {article}  

\usepackage{graphicx}    
\usepackage{amssymb}     
\usepackage{bm}          

\usepackage{cite}

\newcommand{\finalnewpage}{\newpage}  

\newcommand{\beq}{\begin{equation}}
\newcommand{\beqa}{\begin{eqnarray}}
\newcommand{\coords}{t,{\bf x}}
\newcommand{\densr}{n({\mathbf r})}

\newcommand{\D}{\mathrm{d}}
\newcommand{\eeq}{\end{equation}}
\newcommand{\eeqa}{\end{eqnarray}}

\newcommand{\EHK}{E_{{\rm HK}}}
\newcommand{\Eint}{E_{{\rm int}}}
\newcommand{\Exc}{E_{{\rm xc}}}
\newcommand{\FHK}{F_{{\rm HK}}}
\newcommand{\Fni}{F_{{\rm ni}}}
\newcommand{\fpi}{f_\pi}
\newcommand{\fv}{f_{{\rm v}}}
\newcommand{\gA}{g_{A}}
\newcommand{\grad}{{\mbox{\boldmath$\nabla$}}}
\newcommand{\gs}{g_{{\rm s}}}
\newcommand{\gv}{g_{{\rm v}}}

\newcommand{\isovectorTensorN}{{\wt s}_{\tauvec}}

\newcommand{\kf}{k_{\scriptscriptstyle\rm F}}

\newcommand{\kFermi}[1]{%
      k_{{\scriptscriptstyle \rm F}}^{#1}} 
\newcommand{\mc}[1]{\multicolumn{1}{c}{$\quad #1$}}
\newcommand{\mpi}{m_\pi}
\newcommand{\mv}{m_{{\rm v}}}
\newcommand{\ms}{m_{{\rm s}}}
\newcommand{\Mstar}{M^*}
\newcommand{\Mstarsq}{{M^*}^2}
\newcommand{\Nbar}{{\skew3\overline N}}
\newcommand{\newg}{g_{\rho}}
\newcommand{\pcoords}{t,-{\bf x}}
\newcommand{\psibar}{{\overline \psi}}

\newcommand{\QED}{\protect\lower0.0ex\hbox{\rule{2.2mm}{2.2mm}}}

\newcommand{\rhoB}{\rho_{{\scriptscriptstyle \rm B}}}
\newcommand{\rhoBt}{\wt\rho_{{\scriptscriptstyle \rm B}}}
\newcommand{\rhominus}{\rho_{-}}
\newcommand{\rhoplus}{\rho_{+}}
\newcommand{\rhos}{\rho_{{\scriptstyle \rm s}}}
\newcommand{\rhost}{\wt\rho_{{\scriptstyle \rm s}}}

\newcommand{\rhothree}{\rho_{3}}
\newcommand{\rhothreet}{\wt\rho_{3}}

\newcommand{\smk}{\kern2pt}
\newcommand{\tauvec}{{\mbox{\boldmath$\tau$}}}
\newcommand{\stensor}{{{s}}}
\newcommand{\tensorN}{\wt\stensor}

\newcommand{\varDelta}{\Delta}
\newcommand{\varPhi}{\Phi}
\newcommand{\varLambda}{\Lambda}

\newcommand{\veff}{v_{{\rm eff}}}
\newcommand{\wt}{\widetilde}
\newcommand{\zz}{\phantom{0}}

\begin{document}

\title{Covariant Effective Field Theory for Nuclear Structure\\
        and Nuclear Currents}

\author{Brian D. Serot\thanks{Electronic address:
             \texttt{serot@iucf.indiana.edu}}\\[4pt]
             Department of Physics and Nuclear Theory Center\\
             Indiana University, Bloomington, IN\ \ 47405, USA}
\maketitle

\pagestyle{myheadings} 
\thispagestyle{plain}         
\markboth{Brian D.~Serot}{Covariant Effective Field Theory} 
\setcounter{page}{1}         

\section{Introduction}
\label{sec:INTRO}

The fundamental theory of the strong interaction is quantum
chromodynamics (QCD), which is a relativistic field theory with
local gauge invariance, whose elementary constituents are colored
quarks and gluons.
In principle, QCD should provide a complete description of nuclear 
structure and dynamics.
Unfortunately, QCD predictions at nuclear length scales with the
precision of existing (and anticipated) experimental data are not
available, and this state of affairs will probably persist for
some time.
Even if it becomes possible to use QCD to describe nuclei directly,
this description is likely to be cumbersome and inefficient, since
quarks cluster into hadrons at low energies.

How can we simplify this problem to make progress?
We will employ a framework based on Lorentz-covariant,
effective quantum field theory and density functional theory.
Effective field theory (EFT) embodies basic principles that are 
common to many areas of physics, such as the natural separation 
of length scales in the description of physical phenomena.
In EFT, the long-range dynamics is included explicitly, while
the short-range dynamics is parametrized generically; all of
the dynamics is constrained by the symmetries of the interaction.
When based on a local, Lorentz-invariant Lagrangian (density), 
EFT is \emph{the most general way} to parametrize observables
consistent with the principles of quantum mechanics, special
relativity, unitarity, gauge invariance, cluster decomposition, 
microscopic causality, and the required internal symmetries.

Density functional theory (DFT), which has been widely used in
atomic and condensed-matter physics, allows us to describe the 
nuclear many-body system with a universal energy functional that
depends on nuclear densities and four-vector currents.
In principle, knowledge of the full energy functional allows us
to calculate any observable for the (zero-temperature) many-body
system; moreover, a simplified treatment of the functional
based on quasi-particle orbitals still provides an 
\emph{exact\/} description
of bulk nuclear properties and some single-particle observables.
The great advantage of DFT is that calculations of this subset of
properties can be made without knowledge of the many-body wave
function, or with a simple one.
Finally, if relevant expansion parameters can be found, the energy
functional can be truncated to a manageable size, and the accuracy
of the truncation can be tested quantitatively for the observables
in question.

The basic properties of nuclei provide stringent constraints on any
nuclear theory.
An accurate description of these properties is necessary for any
useful predictions or extrapolations.
We certainly want to reproduce the observed shapes of nuclei: the
interior density of a heavy nucleus should be relatively constant,
there should be a well-defined surface, and because of nuclear
``saturation'', the radius $R$ of a nucleus should scale according
to $R \propto B^{1/3}$, where $B = N + Z$ is the total number of
neutrons and protons.
Moreover, the total energy $E$ of the nucleus should agree with the 
``liquid drop'' formula
\begin{equation}
E = -{a_1}B + {a_2}B^{2/3}+  {a_3}Z^2/B^{1/3}
                  + {a_4}(N - Z)^2/B + \cdots \ ,
\label{eq:liquiddrop}
\end{equation}
where typical values for the $a_i$ coefficients are given in
\cite{FW,subatomic,SW86r}.

The particle spectrum is determined by the qualitative
features of the single-particle potential.
In nonrelativistic (Schr\"odinger) terminology, the central 
potential is midway between a harmonic oscillator and a square
well; this shape determines the ordering of the levels as 
a function of the orbital angular momentum.
(See \cite{FW}, Figs.~57.1 and 57.2.)
In addition, the spin-orbit potential is strong, which is
instrumental in determining the major shell closures and, hence,
the nuclear shell model.
We will see later how these features are easily
reproduced in a description based on the Dirac equation.

These simple nuclear features are the ones we will focus on.
We expect that they can be adequately described by a single-particle
equation with an effective, one-body interaction.
Such an approach has many names, depending on the system being
studied and on the practitioner: ``shell model'', ``mean-field
theory'', ``Kohn--Sham'' DFT, etc.
Our goal is to correlate (fit) a modest number of nuclear bulk and 
single-particle data and then to predict other, similar data as
well as possible.

\subsection{Why Use Hadrons?}
\label{sec:hadrons}

Well, why not?
Our focus is on low-energy, long-range nuclear characteristics, 
and all measured observables are colorless.
(In fact, most of the observables of interest to us are dominated by
the \emph{isoscalar} part of the interaction.)
Moreover, hadronic variables (baryons and mesons) are efficient, 
since hadrons are the particles that are observed in experiments.
Colored quarks and gluons participate 
\emph{only in intermediate states},
and such ``off-shell behavior'' is unobservable; by using hadrons, we
expend no theoretical effort combining quarks and gluons into color
singlets that can actually be observed.

So we pick the most efficient degrees of freedom by choosing hadrons.
We will have to parametrize the nuclear EFT Lagrangian anyway, since
we cannot compute its true form from QCD, and hadronic variables, 
if combined in all forms consistent with the underlying
symmetries, provide sufficient flexibility for our parametrization.
We cannot guarantee that a single-particle hadronic approach will be
successful in describing all of the observables of interest, 
but we want to see how well we can do.

\subsection{Why Use the Dirac Equation?}
\label{sec:Dirac}

To motivate the Dirac equation as straightforwardly as possible, 
compare the particle spectrum (and fine structure) in a light atom 
with the spectrum in a heavy nucleus.
An example of the former is given in \cite{BjD}, while an early 
example of the latter is given in \cite{Mayer55}, 
which is reproduced in Fig.~57.3 of \cite{FW}.
The most striking result is that it is impossible to draw the atomic
fine structure to scale, since the splittings are roughly 1/10,000
as large as the major-level splittings (at least for the deeply bound
atomic levels).
In contrast, the nuclear spectrum shows that the ``fine'' structure
is really ``gross''; the spin-orbit splittings are as large as
the major-level splittings to within a factor of two!

The implication is that there \emph{must\/}
be some relativistic effects that
are \emph{important\/} in nuclei (unlike light atoms), and thus
it is much more natural to use the Dirac equation to describe the
quasi-particle nucleon wave functions.

\subsection{Quantum Hadrodynamics (QHD)}
\label{sec:QHD}

We will refer to Lorentz-covariant, meson--baryon, effective
field theories of the nuclear many-body problem as
``quantum hadrodynamics'' or QHD 
\cite{SW86r,JDW74,BDS92r,FST97r,SW97r,FSp00r,FSL00r,EvRev00r,SW00r}.
When QHD is applied within the framework of modern EFT and DFT, 
it provides a quantitative
description of bulk nuclear properties and the spin-orbit
force throughout the Periodic Table \cite{SW97r,FRS98r,FSL00r}.
This success arises from the presence of large 
Lorentz scalar and vector mean fields, 
which imply that there are \emph{large relativistic interaction
effects} in nuclei under normal conditions \cite{EvRev00r}.
There is evidence from QCD sum rules that these large fields are
dynamical consequences of the underlying 
chromodynamics \cite{COHEN91r,FURNST92r}.
Moreover, similar relativistic effects are responsible for the 
efficient description of spin observables in medium-energy 
proton--nucleus scattering using the Relativistic Impulse 
Approximation \cite{RIA83}, 
and they are consistent with the major role played by 
scalar and vector meson exchange in modern boson-exchange models 
of the nucleon--nucleon (NN) interaction \cite{MACHLEIDT89}.
All of these features motivate further investigation into the 
application of QHD to the nuclear many-body problem.

\section{Effective Field Theory}
\label{sec:EFT}

A modern discussion of QHD begins by interpreting the Lagrangian
as defining a nonrenormalizable, effective field theory (EFT)
\cite{FTS95,SW97r}.
An effective Lagrangian consists of known long-range interactions
constrained by symmetries and a complete, non-redundant set of 
generic short-range interactions (i.e., ``contact'' and ``gradient'' terms).
The division between ``long'' and ``short'' is characterized by the
\emph{breakdown scale\/} $\varLambda$ of the EFT.
While it is not possible at present to derive an effective hadronic
theory directly from the underlying QCD, the EFT perspective implies
that this is not necessary.
If one constructs a general Lagrangian that respects the symmetries of
QCD: Lorentz covariance, parity conservation, time-reversal 
and charge-conjugation invariance, (approximate) isospin symmetry,
and spontaneously broken chiral symmetry, then the EFT is a general
parametrization of observables below the breakdown scale.\footnote{%
It is straightforward to include the local U(1) gauge symmetry of
the electromagnetic interaction \protect\cite{FST97r,SW97r}.}

For QHD, we identify $\varLambda$ with the mass scale of physics
beyond the Goldstone bosons (pions); we will see that $\varLambda
\approx 600\,\mathrm{MeV}$.
At momenta small compared to $\varLambda$, short-distance physics
(such as the substructure of nucleons) is only partially resolved
and so may be incorporated into the coefficients of field operators
organized as a derivative expansion.
The coefficients of these short-range terms may eventually be derived
from QCD, but at present, they must be fitted by matching calculated
and experimental observables.  In principle, there are an infinite
number of (nonrenormalizable) terms, but in practice, the Lagrangian
or energy functional can be truncated to work to a given precision
\cite{FST97r}.
The EFT is useful if this truncation can be made at low enough order
that the number of free parameters is not prohibitive.

The EFT perspective, with the freedom to redefine and transform fields,
implies that \emph{there are infinitely many representations of
low-energy \emph{QCD} physics}.
But they are not all equally efficient or physically transparent.
One of the possible choices is between Lorentz-covariant and
nonrelativistic formulations.
Recent developments in baryon chiral perturbation theory support
the consistency (and utility) of a covariant EFT, with Dirac nucleon
fields in a Lorentz-invariant, effective Lagrangian density
\cite{TANG96,ELLIS98,BECHER99}.
A similar framework underlies QHD approaches to nuclei.

In QHD, the only \emph{essential\/} degrees of freedom are the 
nucleons and pions. 
Only these stable particles can appear on external lines with
\emph{timelike\/} four-momenta.
The long-range pion--pion and pion--nucleon interactions are included
in a nonlinear realization of the 
spontaneously broken $\mathrm{SU(2)}_L \times \mathrm{SU(2)}_R$ 
chiral symmetry, which avoids dynamical assumptions inherent in
linear representations.
These interactions can be written down systematically, given an
appropriate power-counting scheme, to be discussed shortly
\cite{FST97r}.
Low-mass vector mesons are typically included for phenomenological
reasons, but are not required, since their masses are roughly 
equal to the breakdown scale $\varLambda$; they are absent from
point-coupling Lagrangians, for example
\cite{NIKOLAUS92,RUSNAK97}.
In descriptions of NN scattering and of nuclear structure and reactions,
the heavy, non-Goldstone bosons appear only on internal lines
(with \emph{spacelike\/} four-momenta) and allow us to parametrize
the medium- and short-range parts of the NN interaction, as well as 
the electromagnetic form factors of the hadrons
\cite{FTS95,FST97r}.
The heavy bosons are also convenient degrees of freedom for
describing nonvanishing expectation values of bilinear nucleon
operators, such as $\Nbar N$ and $\Nbar \gamma^\mu N$, which are 
important in nuclear many-body systems \cite{JDW74,SW86r,SW97r}.
This explains why it is useful to introduce collective degrees of
freedom with other quantum numbers, such as a $\varDelta$ baryon to
incorporate important pion--nucleon interactions 
\cite{ELLIS98,DELTA96}.
Because one must always truncate the EFT Lagrangian, these degrees
of freedom can be efficient in the many-body problem \emph{whether
or not they are actually observed as hadronic resonances}.

A Lorentz scalar, isoscalar mean field in nuclei is an efficient
way to include implicitly the effects of pion exchange that are the
most important for describing bulk nuclear properties.
Because the chiral symmetry is realized nonlinearly, one can
\emph{add\/} to the theory a light scalar, isoscalar, chiral-singlet 
field with a Yukawa coupling to the nucleon, just as in
the original Walecka model \cite{JDW74}.
Nonlinear self-interactions of this new scalar must be included, 
with adjustable couplings that arise in part from the nucleon
substructure.
Since the expectation value of the pion field in nuclear matter
vanishes at the mean-field level, one makes the remarkable observation
that the mean-field theory (MFT) of the Walecka model \emph{is
consistent with chiral symmetry}, provided we think in terms of a
nonlinear realization of the symmetry.
The light scalar, isoscalar field, which is \emph{not\/} the chiral
partner of the pion, plays the same role in the EFT as it does in
the Walecka model: it simulates important $\pi\pi$ and NN interactions
that must be included from the outset to generate a realistic
description of nuclear matter and nuclei.

To make systematic calculations, the EFT approach exploits the
separation of scales in physical systems, with the ratios of scales
providing expansion parameters.
A connection between appropriate QCD scales and nuclear phenomenology
is made by applying Georgi and Manohar's naive dimensional analysis
(NDA) \cite{GEORGI84,GEORGI93}
and \emph{naturalness}, namely, that all appropriately defined, 
dimensionless couplings are of order unity.
With this input, the nonlinear chiral Lagrangian can be organized in
increasing powers of the fields and their derivatives.
To each interaction term we assign an index
\begin{equation}
\nu = d + \frac{n}{2} + b \ ,
\label{eq:nudef}
\end{equation}
where $d$ is the number of derivatives, $n$ is the number of nucleon
fields, and $b$ is the number of non-Goldstone boson fields in the
interaction term.
Derivatives on the nucleon fields are not counted in $d$ because they
will typically introduce powers of the nucleon mass $M$, which will
not lead to small expansion parameters.

It was shown in \cite{FTS95,Fur96,FST97r} that for finite-density 
applications at and below nuclear matter equilibrium density, one can 
truncate the effective Lagrangian to terms with $\nu \leq 4$.
It was also argued that by making suitable definitions 
of the nucleon and meson fields, 
it is possible to write the Lagrangian in a ``canonical'' form
containing familiar noninteracting terms for all fields, 
Yukawa couplings between the nucleon and meson fields, 
and nonlinear meson interactions \cite{FHT01}.
See \cite{FST97r,SW97r} for a more complete discussion.

If we keep terms with $\nu \leq 4$, the chirally invariant Lagrangian
can be written as\footnote{%
We use the conventions of \protect\cite{SW86r,FST97r,SW97r}.
The pion-decay constant is 
$\fpi \approx 93\,\mathrm{MeV}$.}${}^,$\footnote{\label{fn:alphas}%
The two terms involving $\alpha_i$ coefficients actually have
$\nu = 5$, but they were found to be numerically significant
in \protect\cite{FST97r}.
See Fig.~\protect\ref{fig:be_pc_o16}, below.}
\begin{eqnarray}
{\cal L}_{\mathrm{EFT}} & \equiv & {\cal L}_N  + {\cal L}_4
        + {\cal L}_M  \nonumber \\[7pt]
 & = & \Nbar \left( i {\gamma}^{\mu} \left[ {\partial}_{\mu} + i v_{\mu}
+ i g_{\rho} {\rho}_{\mu} + i \gv V_{\mu} \right]
+ {\gA}\mkern2mu
  {\gamma}^{\mu} {\gamma}_{5} a_{\mu} - M + \gs \phi \right) N 
        \nonumber \\[4pt]
 & & \quad
{} -  { {f_{\rho} g_{\rho}} \over {4 M} } 
\Nbar {\rho}_{\mu \nu} {\sigma}^{\mu \nu} N 
- { {\fv \gv} \over {4 M} } 
\Nbar \,{V}_{\mu \nu} {\sigma}^{\mu \nu} N
        \nonumber \\[4pt]
 & & \quad 
{} - { { {\kappa}_{\pi} } \over M } 
\Nbar \,{v}_{\mu \nu} {\sigma}^{\mu \nu} N 
+ { {4 {\beta}_{\pi}} \over M } \Nbar N \, {\rm Tr} 
\left( a_{\mu} a^{\mu} \right) 
  + {\cal L}_4
\nonumber \\[4pt]
 & & \quad
{} +  { 1 \over 4 } f^2_{\pi} \,
{\rm Tr} \left({\partial}_{\mu} U {\partial}^{\mu} 
                U^{\dagger} \right)
+  { 1 \over 2 } \,
  \left( 1 + \alpha_1 \frac{\gs \phi}{M} \right) 
{\partial}_{\mu} \phi \, {\partial}^{\mu} \phi 
\nonumber \\[4pt]
 & & \quad
{} - { 1 \over 4 }\, 
   \left( 1 + \alpha_2 \frac{\gs \phi}{M} \right)
{V}_{\mu \nu} {V}^{\mu \nu} 
- { 1 \over 2 } \, {\rm Tr} 
      \left( {\rho}_{\mu \nu} {\rho}^{\mu \nu} \right) 
\nonumber \\[4pt]
 & & \quad
 {} - g_{\rho \pi \pi} { {2 f^2_{\pi}} \over { m^2_{\rho} } } \,
{\rm Tr} \left( {\rho}_{\mu \nu} {v}^{\mu \nu} \right)
+ { 1 \over 2 } \left( 1 + {\eta}_1 { {\gs \phi} \over M } 
+ {{\eta}_2 \over 2} { {g^2_s {\phi}^2} \over {M^2} } \right)
\mv^2 V_{\mu} V^{\mu} 
\nonumber \\[4pt]
 & &\quad
 {} + { 1 \over {4!} } \,{\zeta}_0 \gv^2 
   {\left( V_{\mu} V^{\mu} \right)}^2
+ \left( 1 + {\eta}_{\rho} \, { {\gs \phi} \over M } \right)
m^2_{\rho} \, {\rm Tr} \left( {\rho}_{\mu} {\rho}^{\mu} \right)
\nonumber \\[4pt]
 & &\quad
{} - \ms^2 {\phi}^2 \left( { 1 \over 2 } + { {{\kappa}_3 } \over {3!} }
{ {\gs \phi} \over M } + { {{\kappa}_4 } \over {4!} } 
{ {g^2_s {\phi}^2} \over {M^2} }\right) ,
\label{eq:eft-lagrangian}
\end{eqnarray}
where the nucleon, pion, sigma, omega, and rho fields are denoted by
$N$, $\mbox{\boldmath $ \pi $}$, $\phi$, $V_\mu$, and $\rho_\mu \equiv
{ 1 \over 2 } \, \mbox{\boldmath $ \tau  \! \cdot \! \rho $}_{\mu}$, 
respectively, 
$V_{\mu\nu} \equiv \partial_\mu V_\nu - \partial_\nu V_\mu$, and
${\sigma}^{\mu \nu} \equiv {i \over 2} [{\gamma}^{\mu}, {\gamma}^{\nu}]$.
The trace ``Tr'' is in the $2 \times 2$ isospin space.
The pion field enters through the combinations
\begin{eqnarray}
U & \equiv & \exp(i \mbox{\boldmath $\tau  \! \cdot \! \pi$} / 
        \fpi ) \ , 
\qquad \qquad
 \xi  \equiv  \exp(i \mbox{\boldmath $\tau  \! \cdot \!\pi$} / 
        2 \fpi ) \ , \label{eq:Upi} \\[5pt]
a_{\mu} &\equiv & - {i \over 2} \left( {\xi}^{\dag} {\partial}_{\mu} {\xi} -
\xi {\partial}_{\mu} {\xi}^{\dag} \right) 
= a_{\mu}^{\dagger} \ , \label{eq:a-mu} \\[5pt]
%
%
v_{\mu} &\equiv&  - {i \over 2} \left( {\xi}^{\dag} {\partial}_{\mu} {\xi} +
\xi {\partial}_{\mu} {\xi}^{\dag} \right) 
 = v_{\mu}^{\dagger} \ , \label{eq:v-mu} \\[5pt]
%
%
v_{\mu \nu} &\equiv & \partial_\mu v_\nu - \partial_\nu v_\mu + i[v_\mu , 
v_\nu ] = - i [a_{\mu},a_{\nu}] \ . \label{eq:v-mu-nu} 
%
%
\end{eqnarray}
The rho meson enters through the chirally covariant field tensor
\begin{equation}
{\rho}_{\mu \nu}=D_{\mu} {\rho}_{\nu} - D_{\nu} {\rho}_{\mu} +i \, \newg
[{\rho}_{\mu}, {\rho}_{\nu}] \ ,
\label{eq:rhofieldtensor}
\end{equation}
where the covariant derivative is defined by
\begin{equation}
D_{\mu} {\rho}_{\nu} \equiv {\partial}_{\mu}{\rho}_{\nu} + i [v_{\mu},
{\rho}_{\nu}] \ .
\label{eq:rhoderiv}
\end{equation}

The antisymmetric combination of derivatives in ${\rho}_{\mu \nu}$
implies that the timelike components $\rho^a_0$ 
of the rho field have no conjugate momenta and are
thus determined by equations of constraint, 
as appropriate for a massive vector field with three dynamical 
degrees of freedom.
The final term in (\ref{eq:rhofieldtensor})
has the usual form for a nonabelian vector field and enables the
rho meson to couple to a conserved isovector current 
\cite{SW86r,BDS79}.
${\cal L}_4$ contains $\pi\pi$ and $\pi$N interactions of order
$\nu = 4$ that are not needed in this work.
Numerically insignificant $\nu = 4$ terms proportional to
$\phi^2 \,{\rm Tr} \left( {\rho}_{\mu} {\rho}^{\mu} \right)$ 
and $V_\nu V^\nu \,{\rm Tr} \left( {\rho}_{\mu} {\rho}^{\mu} \right)$
have been omitted.

To exhibit the chiral invariance of ${\cal L}_{\mathrm{EFT}}$ 
explicitly, we follow CCWZ \cite{CWZ,CCWZ,SW97r}.
A nonlinear realization of the chiral group 
$\mathrm{SU(2)}_L \times \mathrm{SU(2)}_R$ is defined such
that for arbitrary \emph{global\/} matrices $L \in \mathrm{SU(2)}_L$
and $R \in \mathrm{SU(2)}_R$, there is a mapping
\beq
L \otimes R:\quad (\xi, \rho_\mu, N)\longrightarrow 
        (\xi', \rho'_\mu, N')   
          \ .     \label{eq:nonlr}
\eeq
Because of the parity operation ${\cal P}$, 
which produces the transformation
\beq
{\cal P}:\quad L \longleftrightarrow R\ , \quad
      \pi^a (\coords ) \longrightarrow -\pi^a (\pcoords)\ , \quad
      \xi (\coords ) \longrightarrow \xi^\dagger (\pcoords)\ ,
        \label{eq:piparity}
\eeq
the chiral mapping (\ref{eq:nonlr}) can be written as \cite{CWZ}
\begin{eqnarray}
\xi'(x) &=& L \xi(x) h^{\dagger}(x) = h (x) \xi(x) R^{\dagger}
               \ , \label{eq:Xitrans} \\[4pt]
\rho'_\mu(x) &=& h (x) \rho_\mu (x) h^{\dagger}(x)
               \ , \label{eq:Rhotrans} \\[4pt]
 N'(x) &=& h (x) N(x)  \ .       \label{eq:Ntrans}
\end{eqnarray}
The second equality in (\ref{eq:Xitrans}) defines
$h (x)$ as a function of $L$, $R$, and the local pion fields:
$
h (x)=h (L,R, \mbox{\boldmath $\pi$} (x)).
$
It follows from (\ref{eq:Xitrans}) that $h (x)$ is invariant 
under the parity operation (\ref{eq:piparity}), that is,
\beq
h (x) \in \mathrm{SU(2)}_{V} \ ,
\eeq
where $\mathrm{SU(2)}_{V}$ is the unbroken vector subgroup of 
$\mathrm{SU(2)}_{L} \times \mathrm{SU(2)}_{R}$.
Note that the matrix $h (x)$  becomes a \emph{constant\/} only when 
$L = R$, so that $h=L=R$.
Equations~(\ref{eq:Rhotrans}) and (\ref{eq:Ntrans}) then
ensure that the rho and nucleon fields transform linearly under
global $\mathrm{SU(2)}_{V}$, in accordance with their isospins. 
The isoscalar fields $V_{\mu} (x)$ and $\phi (x)$ are chiral scalars
and are unaffected by both chiral and isospin transformations.

For discussing purely pionic interactions, it is convenient to 
use the matrix $U(x)$ of (\ref{eq:Upi}),
since the transformation law (\ref{eq:Xitrans}) then implies
\beq
U(x) \longrightarrow U' (x) =
  L U (x) R^\dagger \ , \label{eq:Utrans}
\eeq
so that $U (x)$ {\em always transforms globally}.
Thus derivatives of $U (x)$ transform the same way as $U (x)$, and
chirally invariant interactions involving pions alone can be 
constructed from products of $U (x)$, $U^\dagger(x)$, 
and their derivatives.
As is well known, these terms can be organized according to the 
number of derivatives, resulting in the Lagrangian of chiral
perturbation theory \cite{subatomic,BIRA99}.
We will return to this later when we discuss electroweak interactions
with nuclei.

For describing the interactions of pions with other particles, 
$U (x)$ is not convenient, because other fields transform with the 
\emph{local\/} functions $h (x)$ of the unbroken isovector subgroup 
$\mathrm{SU(2)}_{V}$.
It follows from the transformation laws given earlier that 
interaction terms that are invariant under {\em local\/} isospin 
rotations will be invariant under {\em global\/} transformations 
of the full group $\mathrm{SU(2)}_{L} \times \mathrm{SU(2)}_{R}$.
Thus, to form chirally invariant interactions involving pions and 
other fields, we need functions of the pion field that transform 
with $h (x)$ only.

The desired functions involving one derivative of the pion field
are given in (\ref{eq:a-mu}) and (\ref{eq:v-mu}).
The parity transformation (\ref{eq:piparity}) implies that
$a_{\mu}$ is an axial vector and $v_{\mu}$ is a polar vector.
Moreover, under a chiral transformation, (\ref{eq:Xitrans}) implies
\begin{eqnarray}
a_\mu   &\longrightarrow &  a'_\mu =  h a_\mu h^{\dagger} 
               \ , \\[3pt]
v_\mu   &\longrightarrow  & v'_\mu =  h v_\mu h^{\dagger}
                   -i h\partial_\mu h^{\dagger} 
  = h v_\mu h^{\dagger} + i (\partial_\mu h ) h^{\dagger} \ .
\end{eqnarray}
Thus $a_{\mu}$ transforms {\em homogeneously\/} under the local 
$\mathrm{SU(2)}_{V}$ group and can be interpreted as a 
\emph{covariant derivative\/} of the pion-field matrix $\xi (x)$.
In contrast, the {\em inhomogeneous\/} transformation law for 
$v_{\mu}$ resembles that of a gauge field, so that $v_{\mu}$ 
allows us to construct chirally covariant derivatives of the 
other fields.
For example, it is straightforward to verify that the covariant 
derivatives (\ref{eq:rhoderiv}) and
\beq
D_{\mu} N \equiv (\partial_{\mu} + i v_{\mu}) N 
          \label{eq:Ncovderiv}
\eeq
transform homogeneously with $h(x)$ under the full group:
\beq
(D_{\mu} N )' = h (D_{\mu} N ) \ , \quad
(D_\mu \rho_\nu )' = h (D_\mu \rho_\nu ) h^{\dagger} \ .
\eeq

The covariant tensor for the pion field is $v_{\mu\nu}$
[see (\ref{eq:v-mu-nu})],
which transforms homogeneously with $h$, as does 
${\rho}_{\mu \nu}$.
This allows us to produce a chirally invariant $\rho\pi\pi$ coupling 
through an interaction of the form 
${\rm Tr} \, ( \rho_{\mu\nu} v^{\mu\nu} )$.

Electromagnetic interactions can be included by adding a chirally
noninvariant Lagrangian [we exhibit terms of $O(e)$ only]
\finalnewpage      
  \beqa
{\cal L}_{\rm EM} & = & - \frac{1}{4} \, F_{\mu\nu} F^{\mu\nu}
      - 2e \fpi^2 A^\mu \, {\rm Tr}
      \left( v_\mu \tau_3 \right) 
     - \frac{e}{2g_\gamma}\, F_{\mu\nu} \left[
      {\rm Tr} \left( \rho^{\mu\nu} \tau_3 \right) 
     + \textstyle{\frac{1}{3}}
      V^{\mu\nu} \right] \nonumber \\[4pt]
& & \quad
{} -{1\over 2}\, e A^\mu \Nbar 
     (1+\tau_3) \gamma_\mu N
    - {e\over 4M} \, F^{\mu\nu} \, \Nbar
   \left(\lambda^{(0)}+ \lambda^{(1)} \tau_3\right) \sigma_{\mu\nu} N
        \nonumber \\[6pt]
& & \quad {} -{e\over 2M^2} \,
       \partial_\nu F^{\mu\nu} \, \Nbar \left[
	\left(\beta^{(0)} + \beta^{(1)} \tau_3\right)\gamma_\mu \right]
      N \ ,
  \label{eq:EMLagrangian}
 \eeqa
where $A^\mu$ is the photon field and $F_{\mu\nu} \equiv
\partial_\mu A_\nu - \partial_\nu A_\mu$ is the usual Faraday tensor.
The constants $e$, $\lambda^{(t)}$, and $\beta^{(t)}$ in 
(\ref{eq:EMLagrangian}) (where $t = 0, 1$ denotes the isospin), 
together with the tensor couplings $\fv$ and $f_{\rho}$ 
in (\ref{eq:eft-lagrangian}),
are sufficient to parametrize the empirical nucleon charge $e$, 
the anomalous moments $\lambda_\mathrm{p,n}$, and the charge and 
magnetic radii $( r_\mathrm{rms} )^{(t)}_{1,2}$ 
at low momentum transfer \cite{FST97r}.
The expansion can be extended to include higher derivatives of
the photon field if greater accuracy is needed \cite{RUSNAK97}.

Similarly, the free parameter $\gA$ in the pion--nucleon 
interaction [see (\ref{eq:eft-lagrangian})]
allows us to normalize the one-body, axial-vector nuclear 
current so that the Goldberger--Treiman relation is satisfied at
the tree level \cite{AXC}.

To summarize the important points of the full Lagrangian
${\cal L}_{\mathrm{QHD}} \equiv {\cal L}_{\mathrm{EFT}} 
+ {\cal L}_{\rm EM}$
[recall (\ref{eq:eft-lagrangian}) and (\ref{eq:EMLagrangian})]:
\begin{itemize}
\item
The noninteracting hadron terms take their standard canonical forms.
\item
The generalized coordinates (fields) have been chosen so that the
meson--nucleon couplings have a simple Yukawa form.
\item
The pion--nucleon and pion--meson interactions enforce 
the nonlinear realization of chiral symmetry.
\item
The nonlinearities involving chiral singlet fields are obviously
invariant, and fitting their coefficients to data will implicitly
include short-range dynamics from many-nucleon forces, fluctuations
of the quantum vacuum, and hadron substructure.
\item
The nucleon electromagnetic (and weak) structure $(\gA, \lambda,
\mathrm{etc.})$ is included to the desired accuracy using a
derivative expansion of the fields.
\end{itemize}

\section{Density Functional Theory}
\label{sec:DFT}

The successes of QHD mean-field phenomenology are, at first, rather
mysterious from the EFT perspective alone, since the Hartree
approximation is just the finite-density counterpart of the Born
approximation at zero density.
The density functional theory (DFT) perspective explains the
successes of mean-field approaches and provides a new context for
EFT power counting.

We begin with a discussion of nonrelativistic DFT and generalize 
later to include relativity.
The basic idea behind DFT is to compute the energy $E$ of the
many-fermion system (or, at finite temperature, the grand
potential $\Omega$) as a functional of the particle 
density \cite{DREIZLER90}.
DFT is therefore a successor to Thomas--Fermi theory 
\cite{T27,F28}, which uses a crude energy functional, 
but eliminates the need to calculate the many-fermion wave function.

The strategy behind DFT can be seen most easily by working
in analogy to thermodynamics \cite{thermo}. 
For a uniform system in a box of volume $V$ at temperature $T$,
one first computes the grand potential $\Omega (\mu , T, V)$, where
$\mu$ is the chemical potential.
It then follows that the number of particles $N$ is determined by
\cite{FW}
\begin{equation}
N = \langle {\widehat N} \mkern2mu \rangle = {} - \left(
       \frac{\partial \Omega}{\partial \mu} \right)_{T,V} \ .
  \label{eq:number}
\end{equation}
According to Gibbs, thermodynamic equilibrium is defined by the
condition
\beq
 (\delta \Omega )_{\mu , T, V} \geq 0 \ ;
\eeq
an assembly minimizes its grand potential at fixed $\mu$,
$V$, and $T$.
Thus the convexity of $\Omega$ implies that $N$ is a monotonically
increasing function of $\mu$, so the relation (\ref{eq:number})
can be inverted for $\mu (N)$.
Finally, one makes a Legendre transformation to the Helmholtz
free energy 
\beq
F(N,T,V) \equiv \Omega (\mu (N),T,V) + \mu(N) N
    \label{eq:Helmholtz}
\eeq
to discuss systems with a fixed density $n \equiv N/V$.

For a self-bound, finite system, 
we replace the chemical potential with an
external, single-particle potential\footnote{%
Here the chemical potential $\mu$ is absorbed into the definition 
of $v$, which defines the zero of energy.  
We suppress all spin and isospin dependence at this point.}
$\sum_i v({\mathbf r}_i)$.
The grand potential is now a \emph{functional}:
$\Omega [v ({\mathbf r}); T)$, and a functional derivative
with respect to $v$ gives the particle density:\footnote{%
Higher variational derivatives yield various correlation 
functions.}
\begin{equation}
n ({\mathbf r}) = \langle {\hat n}({\mathbf r}) \rangle 
= \frac{\delta \Omega}{\delta v ({\mathbf r})}  \ .
   \label{eq:ndens}
\end{equation}
The convexity of $\Omega$ allows us to invert this
relation (in principle) and to find $v ({\mathbf r})$ 
as a (complicated) functional of $n({\mathbf r})$:
\begin{equation}
v ({\mathbf r}) = v[ n({\mathbf r}) ] \ .
\end{equation}
($T$ is suppressed.)
Thus there is a one-to-one relation between the external potential
and the particle density.
Needless to say, the possibility of this complicated inversion is
a matter of great technical interest, and the reader is referred to
\cite{DREIZLER90} for a discussion.

Finally, we make a functional Legendre transformation to define the
Hohenberg--Kohn free energy, which is a functional of
$n({\mathbf r})$:
\begin{equation}
\FHK [n({\mathbf r})] = \Omega 
       \left[v[ \, n({\mathbf r})]\, \right]
    - \int {\mathrm d}{\mathbf r} \ n({\mathbf r}) v({\mathbf r})
\ .
   \label{eq:FHK}
\end{equation}
The variational derivative of this free-energy functional with
respect to $n$ now gives
\begin{equation}
 \frac{\delta \FHK [n]}{\delta n({\mathbf r})}
   = - v({\mathbf r}) \ ,
    \label{eq:varF}
\end{equation}
where we have used (\ref{eq:ndens}).

If we now restrict consideration to $T=0$ and $v({\mathbf r}) =0$,
the free energy becomes simply the energy:
$\FHK [n] \longrightarrow \EHK [n]$, and
the {\em Hohenberg--Kohn theorem} follows \cite{SW97r,KOHN99}:
If the functional form of $\EHK [n({\mathbf r})]$ is
known exactly, the ground-state expectation value of any
observable is a \emph{unique\/} functional of the exact 
ground-state density.
Moreover, it follows immediately from (\ref{eq:varF}) that
the exact ground-state density can be found by minimizing the
energy functional.
Although we have assumed here that the ground state is 
non-degenerate, this assumption can be easily relaxed \cite{KOHN99}.

Significant progress in solving these equations was made by
Kohn and Sham \cite{KS}, who introduced a complete set of
single-particle wave functions.
The exact Hohenberg--Kohn free energy for an inhomogeneous 
(finite) many-body system in an external potential 
takes the form
\beq
   \FHK[\densr] = \Fni[\densr]  + \Eint[\densr] \ ,
   \label{eq:FKS}
\eeq
where the subscripts ``ni'' and ``int'' denote noninteracting
and interacting, respectively.
$\Fni [\densr]$ represents the kinetic energy contribution.
The interaction energy $\Eint [\densr]$ is some functional of the
density (and its derivatives); in the many-body problem, it
contains a Hartree term, an exchange-correlation (``xc'')
contribution, etc. \cite{FW}:
\begin{equation}
\Eint [n] = E_{\mathrm{Hartree}} [n] + \Exc [n] + \cdots
\ .
     \label{eq:eintofn}
\end{equation}
Note that $\Exc$ is generally a \emph{nonlocal\/} and 
\emph{nonanalytic\/} functional
of the density that contains both many-body and short-distance
physics, including vacuum fluctuations and hadron substructure.

Now consider the (nonrelativistic) Schr\"odinger equation in a
potential $\veff ({\mathbf r})$, \emph{which is designed to give
the exact density\/} $\densr$:
\beqa
  \Bigl( -\frac{\hbar^2}{2m}\nabla^2 + v_{\rm eff}({\mathbf r}) 
  \Bigr) \psi_i({\mathbf r}) & = & \epsilon_i \psi_i({\mathbf r})
   \ , \label{eq:KSschrod} \\[4pt]
  n({\mathbf r}) & = & \sum_{i\, =\, 1}^N |\psi_i({\mathbf r})|^2 \ .
\eeqa
In this problem, $\veff$ plays exactly the same role as the
previous $v$, and thus the Hohenberg--Kohn equation (\ref{eq:varF})
gives
\beq
 \frac{\delta \Fni [n]}{\delta n({\mathbf r})}
   = - \veff ({\mathbf r}) \ .
    \label{eq:varFni}
\eeq
By taking the variational derivative of (\ref{eq:FKS}) with
respect to $\densr$, using (\ref{eq:varF}), and rearranging
terms, we find
\beq
\veff ({\mathbf r}) = v({\mathbf r})
   + \frac{\delta \Eint [n]}{\delta \densr}
  \ .
   \label{eq:vintdef}
\eeq
Upon setting the external $v({\mathbf r}) = 0$, 
we obtain the effective potential to be used in the 
Kohn--Sham (KS) equations (\ref{eq:KSschrod}):
\beq
\veff ({\mathbf r}) = \frac{\delta \Eint [n]}{\delta \densr} \ .
   \label{eq:KSveff}
\eeq

Thus, if we know the exact interacting energy as a functional of
the density, we can reproduce the exact interacting density using
a set of \emph{single-particle wave functions}.
Kohn calls these the ``density-optimal'' single-particle
wave functions \cite{KOHN99} (as opposed to Hartree--Fock wave
functions, which are ``total-energy optimal'').

The generalization of DFT to relativistic systems is
straightforward \cite{Mac79}.
The energy $\EHK$ now becomes a functional
of \emph{both\/} the ground-state scalar density $\rhos$ and
the baryon four-current density $B_{\mu}$.
Extremization of the functional gives rise to variational
equations that determine the ground-state densities $\rhos$ and
$\rhoB = B_0$.

These equations can again be simplified by following the Kohn--Sham
approach.
In the relativistic case, the complete set of single-particle wave
functions allows us to recast the variational
equations as Dirac equations for occupied orbitals.
The single-particle Hamiltonian contains {\em local}, 
density-dependent, Lorentz scalar and vector potentials, 
even when the exact energy functional is used.
Moreover, one can introduce auxiliary (scalar and vector) fields
$\varPhi ({\mathbf r})$ and $W ({\mathbf r})$, which
correspond to the local potentials and can therefore be identified
as relativistic KS potentials.
The auxiliary fields are determined by extremizing the energy
functional, which gives rise to a Dirac single-particle Hamiltonian.
The isoscalar part (for spherical nuclei) looks like
\beq
 h_0 = -i \mbox{\boldmath$\alpha \cdot \nabla$} + \beta [ M - \varPhi (r) ]
      + W(r) \ ,
\eeq
where $M$ is the nucleon mass and we define $\Mstar \equiv M -
\varPhi$.
The resulting coupled differential
equations resemble those in a relativistic MFT 
calculation \cite{FST97r,SW97r}.
Note that $\varPhi$ need not be simply proportional to the isoscalar,
scalar field $\phi$.
In fact, $\varPhi$ could be proportional to $\phi$ (as in the
Walecka model), or could be expressed as a sum of scalar and vector
densities (as in relativistic point-coupling theories), or could be
a nonlinear function of $\phi$ (as in modern chiral EFT's).

\finalnewpage  
The strength of the KS approach rests on the following 
theorem:
\begin{quote}
\textsl{
The exact ground-state scalar and vector densities, energy, and
chemical potential for the fully interacting many-fermion system
can be reproduced by a collection of (quasi)fermions moving in
appropriately defined, self-consistent, local, classical fields.
}
\end{quote}
\noindent
The proof is again straightforward \cite{KOHN99}.
Start with a collection of noninteracting fermions moving in 
an externally specified, local, one-body potential.
The exact ground state for this system is known: just calculate
the lowest-energy orbitals and fill them up.\footnote{%
For simplicity, we assume that 
the least-bound orbital is completely
filled, so the ground state is non-degenerate.}
Therefore, if one can find a suitable local, one-body potential
based on an {\em exact} energy functional, the exact
ground state of that system can be determined.
But this potential is precisely the $\veff({\mathbf r})$ 
discussed above, 
which is obtained by differentiating the interaction parts of
$\FHK$ with respect to the various densities.
The resulting one-body potential will generally be density
dependent and thus must be determined self-consistently.

Several points are noteworthy.
Since the single-particle basis constructed
as described above is again ``density optimal'', 
the {\em exact} scalar and vector densities are given by
sums over the squares of the Dirac wave functions, with unit
occupation probability.
Moreover, since these densities are guaranteed to make the
energy functional stationary [the external $v({\mathbf r})
= 0$], the exact ground-state energy is also obtained.
The proof that the eigenvalue of the least-bound state is
exactly the Fermi energy is given in \cite{eF}.
Note, however, that aside from this association, the exact
Kohn--Sham wave functions (and remaining eigenvalues) have no
known, directly observable meaning.

If one knows the exact functional form of
the energy on the densities, one can describe the observables
noted in the theorem exactly (and easily) in terms of the 
Kohn--Sham basis.
Observables of this type are typically the ones calculated in
relativistic MFT's  \cite{HS81,RING96,FST97r}.
Moreover, it has been known for many years \cite{SW86r,RBBG} that
the mean-field contributions dominate the single-particle
potentials at ordinary densities.
Thus, by {\em parametrizing} the energy functional in a mean-field
(or ``factorized'') form, and by fitting the parameters to
empirical bulk and single-particle nuclear data, one
should obtain an excellent approximation to the exact energy
functional in the relevant density regime.
This is the key to the success of relativistic MFT calculations,
as we will verify below, using the effective chiral Lagrangian
constructed in Sect.~\ref{sec:EFT}.

\section{Naive Dimensional Analysis}
\label{sec:NDA}

There is still an important point to be addressed:
we must understand how to extract the dimensional scales of 
each term in the Lagrangian, so that the remaining dimensionless 
constants can be checked for naturalness.
A naive dimensional analysis (NDA) for assigning a coefficient of 
the appropriate size to any term in the effective Lagrangian has 
been proposed by Georgi and Manohar \cite{GEORGI84,GEORGI93}.
This allows for a determination of both the dimensional scales 
associated with each term and for the inclusion of an overall 
dimensionless constant that can be used to adjust the strength.
The basic idea of naturalness is that once the appropriate
dimensional scales have been extracted, the overall dimensionless
coefficients should all be of order unity.

The NDA rules for a given term in the Lagrangian density are
\cite{GEORGI93}:
\begin{enumerate}
\item Include a factor of $1/\fpi$ for each strongly interacting
      field.
\item Assign an overall factor of $\fpi^2 \varLambda^2$.
\item Multiply by factors of $1/ \varLambda$ to achieve dimension 
      (mass)$^4$.
\item Include appropriate counting factors (such as $1/n!$ for
      $\phi^n$).
\end{enumerate}
Here $\fpi \approx 93\,\mathrm{MeV}$ is the pion-decay constant, 
and the breakdown scale $\varLambda \approx 600\,\mathrm{MeV}$ 
is taken as the generic large-momentum-cutoff scale, which 
characterizes the mass scale of physics beyond the Goldstone bosons.

As noted by Georgi \cite{GEORGI93}, rule (1) simply assumes that the
amplitude for producing any strongly interacting particle is 
proportional to the amplitude $\fpi$ for emitting a Goldstone boson.
This is a reasonable assumption, since $\fpi$ is the only natural 
scale.
Thus, by dividing each field by $\fpi$, we should arrive at a factor 
of $O(1)$.
Rule (2) can be understood as an overall normalization factor that 
arises from the standard way of writing the mass terms of 
non-Goldstone bosons. 
For example, one may write the mass term of a isoscalar, scalar
field $\phi(x)$ as
\beq
     {1\over 2}\, \ms^2 \phi^2 ={1\over 2} \, \fpi^2 \varLambda^2 
         {\ms^2 \over \varLambda^2} {\phi^2 \over \fpi^2} \ , 
\eeq
where the scalar mass $\ms$ is treated as roughly the same
size as $\varLambda$.
By applying rule (1) and extracting the overall factor of 
$\fpi^2 \varLambda^2$, the remaining ratios are of $O(1)$.
Since all terms will have the same overall scale factor 
$\fpi^2 \varLambda^2$, higher-order terms  
or terms with gradients of fields will be suppressed by
powers of $1/ \varLambda$ relative to the leading mass terms, 
as a result of ``integrating out'' physics above the scale 
$\varLambda$.
(A simple example is the low-momentum expansion of a tree-level 
propagator for a heavy meson of mass $m_H$, which leads to terms 
with powers of $\partial^2 / m^2_H$.)
It is precisely because of these $1/ \varLambda$ suppression factors
and dimensional analysis that one arrives at rule (3).
The origin of the combinatorial factors in rule (4) is discussed
in \cite{FST97r}.

Applying these rules to a generic term in an effective Lagrangian 
involving the isoscalar fields and the nucleon field leads to 
(the generalization to include the pion, rho, and photon is 
straightforward) \cite{Fr96,FST97r}
\beq
   {\cal L} \sim C \, \frac{1}{m!}\,\frac{1}{n!}\,
     \left( {\psibar\Gamma\psi\over\fpi^2 \varLambda}
        \right)^{\mkern-2mu \ell} 
      \left( {\phi\over\fpi} \right)^{\mkern-2mu m}
      \left( {V\over\fpi} \right)^{\mkern-2mu n}
      \left( {\partial\ \mathrm{or}\ \mpi \over \varLambda}
        \right)^{\mkern-2mu p} 
      \fpi^2 \varLambda^2  \ , 
           \label{eq:NDAgen}
\eeq
where $\psi$ is a baryon field, $\Gamma$ is any Dirac matrix, 
derivatives are denoted generically by $\partial$,
and we have allowed for the possibility of 
chiral-symmetry-violating terms that contain the small parameter 
$\mpi / \varLambda$.
The product of all the dimensional factors then sets the scale in
terms of the pion-decay constant $\fpi$ and the EFT breakdown
scale $\varLambda$.
The overall coupling constant $C$ is dimensionless and of $O(1)$ if
naturalness holds.

These scaling rules imply that a general potential for the 
scalar meson can be expanded as
\beq
      V_{\rm S} = \ms^2 \phi^2 \left( {1\over 2}
              +{\kappa_3\over 3!}\,{\gs \phi \over M}
             + {\kappa_4 \over 4!}\,{\gs^2 \phi^2 \over M^2}
             +\cdots \right) \ ,
      \label{eq:SP}
\eeq
in agreement with the corresponding term in 
(\ref{eq:eft-lagrangian}).
Here we have included a factor of $1/\fpi$ for each power of
$\phi$; these factors are then eliminated in favor of 
$\gs\approx M/\fpi$, which is basically the Goldberger--Treiman 
relation \cite{subatomic}. 
Factorial counting factors are also included, since the NDA rules 
are actually meant to apply to the tree-level scattering amplitude 
generated by the corresponding vertex \cite{We90a,FST97r}.
 
The naturalness hypothesis states that after the dimensional factors
and appropriate combinatorial factors are extracted, the overall 
dimensionless coefficients [$\mkern2mu C$ in (\ref{eq:NDAgen})
and $\kappa_3$, $\kappa_4$, \dots\ in (\ref{eq:SP})]
should be of order unity.
It should be clear, however, that the preceding arguments are 
\emph{not a proof\/} of naturalness, since unknown physical scales 
could generate unnaturally large coefficients.
Moreover, some fitted constants may be unnaturally small, which 
often signals a symmetry of the theory that has not yet been
identified.
These caveats notwithstanding, without the naturalness hypothesis
it is basically impossible to construct an effective Lagrangian 
with any predictive 
power.\footnote{The assumption of renormalizability also leads to
a finite number of parameters and well-defined predictions, but 
does so by imposing \emph{unnatural\/} restrictions on the 
Lagrangian, namely, that many parameters are identically zero in the
absence of relevant symmetry arguments.}
Until one can derive the effective hadronic Lagrangian from QCD, the
naturalness hypothesis must be checked by fitting to 
experimental data, as we will do in the following section.

If truncations of the EFT Lagrangian determined by NDA and 
naturalness are valid, it should also be possible to determine
the level of truncation that exhausts the information content
of the input data.
In other words, we should be able to verify that adding terms
with higher values of $\nu$ does not improve the fits to the
empirical observables of interest.
To this end, it is useful to consider the $\nu = 5$ interactions
\begin{equation}
{\cal L}_5 = 
    - {1\over 5!}\kappa_5 {\gs^3\phi^3\over M^3} \ms^2\phi^2
    + {1\over 3!}\eta_3 {\gs^3\phi^3\over M^3}\cdot
      {1\over 2}\, \mv^2 V_{\mu} V^{\mu}
    + {1\over 4!}\zeta_1 {\gs\phi\over M} \gv^2 (V_\mu V^\mu)^2 \ ,
                      \label{eq:fifth}
\end{equation}
to check that their contributions are either negligible or can
be absorbed into slight modifications of the parameters in the
$\nu \leq 4$ Lagrangian ${\cal L}_{\mathrm{QHD}}$.

\section{Mean-Field Theory of Nuclear Structure}
\label{sec:MFT}

The mean-field equations and energy resulting from
${\cal L}_{\mathrm{QHD}}$ in (\ref{eq:eft-lagrangian}) 
and (\ref{eq:EMLagrangian}) can be derived straightforwardly.
The interested reader is referred to \cite{FST97r}, where the
procedures and results are discussed thoroughly.
One important result is that due to the additional nonrenormalizable 
interactions in ${\cal L}_{\mathrm{EM}}$
between the nucleon and the electromagnetic field, 
and also due to vector-meson dominance, 
the computed nuclear charge density automatically contains the 
effects of nucleon structure, and it is unnecessary to introduce 
an {\em ad hoc\/} form factor.

As we will show in Sect.~\ref{sec:PARAMS}, the full MFT Lagrangian
derived from ${\cal L}_{\mathrm{QHD}}$ has more than enough 
parameters to accurately describe the desired nuclear
properties discussed in Sect.~\ref{sec:INTRO}.
The more important question is whether the parameters fitted to 
nuclei are \emph{natural}.
In \cite{FST97r}, the parameters were determined by calculating a 
set of observables \{$X_{\rm th}^{(i)}$\} for several spherical
nuclei and by adjusting the parameters to minimize a generalized 
$\chi^2$ defined by \cite{NIKOLAUS92}:
\begin{equation}
\chi^2 = \sum_{i} \sum_{X}
         \left[{ X_{\rm exp}^{(i)} - X_{\rm th}^{(i)}
         \over W_{X}^{(i)} X_{\rm exp}^{(i)}} 
         \right]^2 \ ,  
   \label{eq:chisq}
\end{equation}
where $i$ runs over the set of nuclei, $X$ runs over the set of
observables, the subscript ``exp'' indicates the experimental value
of the observable, and $W_{X}^{(i)}$ are the relative weights.
The weights were chosen to be the expected accuracy for the given 
observable in a good fit.
In practice, a reasonable range of weights was tested, 
and the qualitative conclusions discussed below were always 
reproduced.
Some of the considerations relevant to the choice of weights are 
discussed in \cite{FST97r,SW97r}.

The relativistic mean-field equations are solved self-consistently
for the closed-shell nuclei $^{16}$O, $^{40}$Ca, $^{48}$Ca, 
$^{88}$Sr, and $^{208}$Pb, and also in the nuclear matter limit. 
The parameters are then fitted to empirical properties of the charge
densities, the binding energies, and various splittings between
energy levels near the Fermi surface using the figure of merit
$(\chi^2 )$ in (\ref{eq:chisq}).
The full set of $\{ X^{(i)} \}$ comprised 29 calculated and 
empirical values.
When working at the highest order of truncation (essentially
$\nu = 4$), the calculated results are very accurate, as we 
illustrate shortly, but they are too numerous to reproduce here
\cite{FST97r,RUSNAK97,FSp00r}.
Some of the fitted parameters at various levels of truncation
are given in Table~\ref{tab:Gparams}.

\begin{table}
\centering
\caption{Parameter sets from fits to finite nuclei, 
         as described in the text.
         The parameters in the lower portion of the table are
         fitted to the (free) nucleon charge and magnetic form
         factors, and the proton charge $e$ and the anomalous
         moments $\lambda_{\mathrm{p}}$ and $\lambda_{\mathrm{n}}$
         are given their empirical values \protect\cite{FST97r}.}
\vspace{.1in}
\begin{tabular*}{5.505in}{ccrrrrrr}
\hline\hline\\[-4pt]
         & $\nu$ & \mc{W1} & \mc{C1} & \mc{Q1} & \mc{Q2} 
         & \mc{G1} & \mc{G2} \\[3pt]
   \hline\\[-5pt]
$m_{\rm s}/M$    & 2 
  & 0.60305 & 0.53874 & 0.53735 & 0.54268 & 0.53963 & 0.55410   \\     
$g_{\rm s}/4\pi$ & 2 
  & 0.93797 & 0.77756 & 0.81024 & 0.78661 & 0.78532 & 0.83522 \\
$\gv/4\pi$   & 2 
  & 1.13652 & 0.98486 & 1.02125 & 0.97202 & 0.96512 & 1.01560   \\
$g_\rho/4\pi$   & 2 
  & 0.77787 & 0.65053 & 0.70261 & 0.68096 & 0.69844 & 0.75467   \\
$\eta_1$        & 3 
  &         & 0.29577 &         &         & 0.07060    & 0.64992 \\
$\kappa_3$      & 3 
  &         & 1.6698\zz  & 1.6582\zz  & 1.7424\zz  & 2.2067\zz  
 & 3.2467\zz \\     
$\eta_\rho$     & 3 
  &         &         &         &        & $-$0.2722\zz  
  & 0.3901\zz    \\     
$\eta_2$        & 4 
  &         &         &         &         & $-$0.96161 & 0.10975   \\
$\kappa_4$      & 4 
  &         &     & $-$6.6045\zz & $-$8.4836\zz & $-$10.090\zz\zz  
  & 0.63152  \\     
$\zeta_0$       & 4 
  &         &         &         & $-$1.7750\zz & 3.5249\zz     
  & 2.6416\zz    \\     
$\alpha_1$      & 5 
  &         &         &         &        & 1.8549\zz     
& 1.7234\zz    \\     
  $\alpha_2$    & 5
  &         &         &         &        & 1.7880\zz     
  & $-$1.5798\zz \\     
$\fv /4$        &  3
  &         &         &         &        & 0.1079\zz     
  & 0.1734\zz    \\[3pt]
\hline\\[-10pt]     
$f_{\rho}/4$    &  3
  & 0.9332\zz & 1.1159\zz & 1.0332\zz & 1.0660\zz & 1.0393\zz     
  & 0.9619\zz    \\     
$\beta^{(0)}$ &  4
  & $-$0.38482 & $-$0.01915 & $-$0.10689 & 0.01181 
  & 0.02844 & $-$0.09328 \\     
$\beta^{(1)}$ &  4
  & $-$0.54618 & $-$0.07120 & $-$0.26545 & $-$0.18470 
  & $-$0.24992    & $-$0.45964     
 \\[4pt] 
 \hline\hline
\end{tabular*}
\label{tab:Gparams}
\end{table}

\finalnewpage 
To simplify the initial discussion, we restrict consideration to
infinite nuclear matter.
For symmetric matter $(N = Z)$, the energy density through order
$\nu = 4$ is given by \cite{FST97r}
\begin{eqnarray}
     {\cal E}_{\rm MFT} [\Phi, W; \rhoB ]
        & = & W\rhoB + {4\over (2\pi)^3}\! \int_0^{\kf}
       {\kern-.1em}{\rm d}^3{\kern-.1em}{k} \,
           \sqrt{{\bf k}^2+\Mstarsq}
      \nonumber \\[4pt]
     & & \quad
       {} + {1\over \gs^2}\left( {1\over 2} +
         {\kappa_3\over 3!}{\Phi \over M}
       + {\kappa_4\over 4!}{\Phi^2\over M^2} \right) \ms^2 \Phi^2
        \nonumber\\[4pt]
     & & \quad
   {} - {1\over 2\gv^2}
      \left( 1 + \eta_1 {\Phi \over M} 
       + {\eta_2\over 2} {\Phi^2\over M^2} \right) \mv^2 W^2
       - {1\over 4! \gv^2} \zeta_0 W^4  , \qquad
   \label{eq:vmdnm}
\end{eqnarray}
where $\kf$ is the Fermi wavenumber, 
$\rhoB \equiv 2 \kf^3 / 3 \pi^2$, and 
$\Phi \equiv \gs \phi_0 = M - \Mstar$ and $W \equiv \gv V_0$ are 
the scaled fields defined earlier in terms of the scalar and vector
mean fields $\phi_0$ and $V_0$.
For readers who are familiar with the corresponding result in
the Walecka model (Eq.~(3.53) in \cite{SW86r}), one sees that the
MFT nuclear matter energy has been generalized to include additional
nonlinearities that are not allowed in the renormalizable Walecka
model.
The fields $\Phi$ and $W$ are determined by extremization of
${\cal E}$.

What causes the nuclear matter saturation and the relatively small binding
energy?
Let's expand $\mathcal{E}_{\rm MFT} / \rhoB$ from (\ref{eq:vmdnm}) 
in powers of $\kf$ \cite{SW97r} and suppress the nonlinear meson terms
for clarity:
\begin{eqnarray}
\mathcal{E}_{\rm MFT} / \rhoB & = & M + 
    \left[ \frac{3 \kFermi{2}}{10 M}
          -\frac{3 \kFermi{4}}{56 M^3}
          +\frac{  \kFermi{6}}{48 M^5} - \cdots \right] 
          + \frac{\gv^2}{2 \mv^2} \, \rhoB 
          - \frac{\gs^2}{2 \ms^2} \, \rhoB \nonumber\\[3pt]
& &  \quad
     {} + \frac{\gs^2}{\ms^2} \, \frac{\rhoB}{M}
       \left[ \frac{3  \kFermi{2}}{10 M}
        -\frac{36 \kFermi{4}}{175 M^3} + \cdots \right] 
     + \left( \frac{\gs^2 \rhoB}{\ms^2 M} \right)^{\mkern-3mu 2}
       \left[ \frac{3 \kFermi{2}}{10 M} - \cdots \right]
     \nonumber\\[3pt]
& &  \quad
     {} + \left( \frac{\gs^2 \rhoB}{\ms^2 M} \right)^{\mkern-3mu 3}
       \left[ \frac{3 \kFermi{2}}{10 M} - \cdots \right] 
     + \cdots 
\label{eq:expansion}
\end{eqnarray}
The lowest-order Lorentz scalar and vector contributions (which are
proportional to $\rhoB$) \emph{set the scale\/} for the large
mean fields $\Phi$ and $W$.
This scale is consistent with chiral QCD counting rules
\cite{FST97r,FSp00r},
but these two terms \emph{cancel almost exactly\/} in the binding energy,
leading to an anomalously small remainder.
However, they \emph{add constructively\/} in the spin-orbit interaction,
leading to appropriately large spin-orbit splittings in nuclei
\cite{FURRY36,HS81,FRS98r}.

It is important to notice the \emph{different behavior\/} of the vector
and scalar interaction terms in (\ref{eq:expansion}).
Whereas the vector interaction enters at only linear order in $\rhoB$,
the scalar interaction enters at all orders; moreover, the leading scalar
term at every order in $\rhoB$ looks exactly the same, and they all add
constructively.
These terms are precisely what one gets by shifting the nucleon mass in
the nonrelativistic kinetic energy term $3 \kFermi{2}/10 M$ 
from $M \rightarrow \Mstar \approx M - \gs^2 \rhoB / \ms^2$.
These additional, repulsive, velocity-dependent interactions reduce the
strength of the lowest-order, attractive scalar contribution and are crucial
for establishing the location of the equilibrium point of nuclear matter.
Thus the different behavior of the vector and scalar interactions leads
to \emph{large relativistic interaction effects\/}
in the nuclear matter energy density.
In contrast, the relativistic corrections to the kinetic energy (the 
nonleading terms in the first pair of square brackets) are indeed small; 
this is \emph{not\/} where the important ``relativity'' is.

\begin{figure}
 \centering 
 \includegraphics*[width=3.2in]{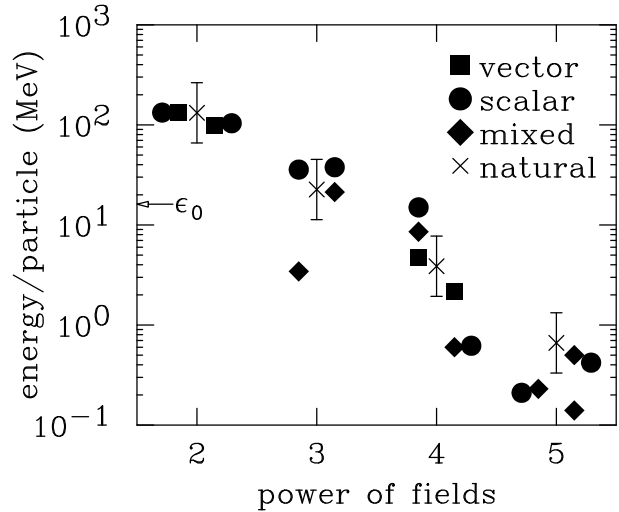}
 \caption{Nuclear matter energy/particle for two QHD parameter sets,
          one on the left (G1) and one on the right (G2) of the
          error bars.
          The power of fields is $b \equiv j + \ell$ for a term of
          the form $\Phi^{\mkern2mu j} W^{\ell}$ ($\ell$ is even).
          The boxes denote terms with $j = 0$, 
          the circles denote terms with $\ell = 0$, and
          absolute values are shown.
          The crosses with error bars are estimates based on 
          \protect(\ref{eq:NDAgen}), with $1/2 \leq C \leq 2$.
          The arrow indicates the total binding energy 
          $\epsilon_0 = 16.0\,\mathrm{MeV}$.
}
 \label{fig:nmbe}
\end{figure}
%
%
\begin{figure}
 \centering 
 \includegraphics*[width=3.2in]{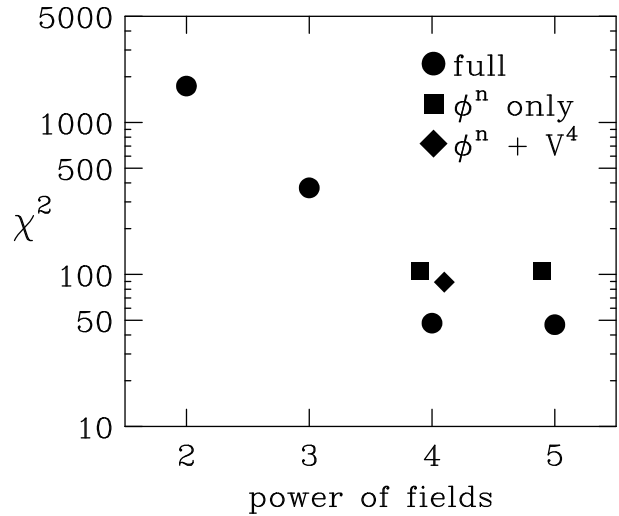}
 \caption{$\chi^2$ values for QHD parameter sets, as a function of
          the level of truncation.
          The power $\nu = 2$ corresponds to set W1, $\nu = 3$ is for
          set C1, the $\nu = 4$ square is for set Q1, the $\nu = 4$
          diamond is for set Q2, and the circle is for the full
          set of parameters G1.
          The $\nu = 5$ results include the terms from ${\cal L}_5$ in
          (\ref{eq:fifth}).
}
 \label{fig:chisq}
\end{figure}

The critical question is whether the hierarchal organization
of interaction terms is actually observed.
This is illustrated in Fig.~\ref{fig:nmbe},
where the nuclear matter energy/particle is shown as a function
of the power of the mean fields, which is called $b$ in
(\ref{eq:nudef}).
(There are no gradient contributions in nuclear matter and
$\langle \hat{\mbox{\boldmath$\pi$}} \rangle = 0$.)
The crosses and error bars are estimates based on NDA and
naturalness, that is, overall coefficients are of order unity.
It is clear that each successive term in the hierarchy is reduced
by roughly a factor of five, which implies a value of
$\varLambda \approx 600\,\mathrm{MeV}$.
Thus for any reasonable desired accuracy, the Lagrangian can be 
truncated at a low value of $\nu$.
In fact, contributions to the energy/particle that are smaller than roughly
$1\,$MeV are below the level of resolution and can be eliminated
in favor of small adjustments of the remaining parameters.
Derivative terms and other coupling terms will be discussed later.

The quality of the fits to finite nuclei and the appropriate level 
of truncation
is illustrated in Fig.~\ref{fig:chisq} \cite{FSp00r}, where the figure of 
merit is plotted as a function of truncation order and of
various combinations of terms retained in ${\cal L}_{\mathrm{QHD}}$.
The full calculations ({\LARGE \lower0.25ex\hbox{$\bullet$}}) 
retain all allowed terms at a given level of $\nu$, 
while the other two choices keep only the indicated subset.
There is clearly a great improvement in the fit (more than a
factor of 35) in going from $\nu = 2$ to $\nu = 4$, but there is
no further improvement in going to $\nu = 5$, using the extra
interactions contained in (\ref{eq:fifth}).
Speaking chronologically, the $\nu = 2$ results show the level
of accuracy obtained more than 20 years ago \cite{HS81}, 
while the $\nu = 4$ results were obtained seven years ago
\cite{FST97r}.
Moreover, the ``$\phi^n$ only'' results at $\nu = 4$ ($\,$\QED$\,$) 
show the state of the situation in the late 1980s, as discussed in
\cite{FPW87,BDS92r}.
Recent work \cite{FSp00r} shows that the full complement of
parameters at order $\nu = 4$ is \emph{underdetermined}, and that only
six or seven are determined by this data set, which explains
the success of the earlier MFT's with a restricted
set of parameters \cite{RING96}.
We will review this analysis in the next section.

\begin{figure}
 \centering 
 \includegraphics*[width=3.45in]{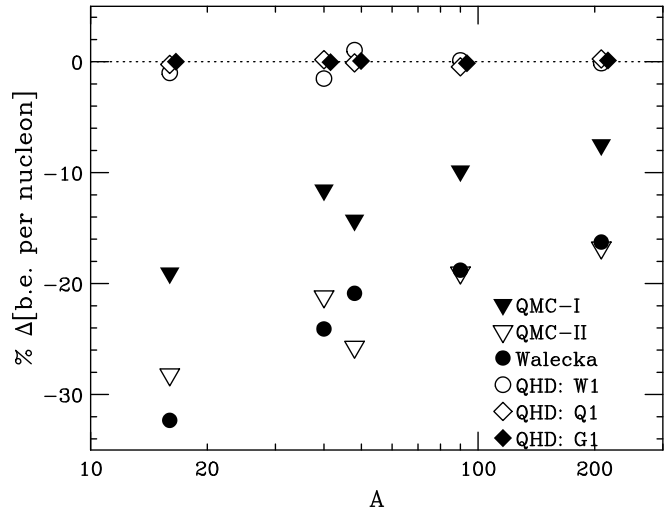}
 \caption{The deviations (in percent) between the calculated binding
          energy/nucleon and the empirical values for five
          doubly-magic nuclei.
          The different calculations and parameter sets are
          discussed in the text.
          For orientation, the Walecka-model results are
          \emph{under}bound.
}
 \label{fig:beG1}
\end{figure}

As a further example of the quality of the fits, and for some
additional historical perspective, Fig.~\ref{fig:beG1}
shows the percent deviation between the calculated and empirical 
binding energies for the five closed-shell nuclei listed earlier.
The results labeled ``Walecka'' are from \cite{HS81}, those
labeled ``QMC'' are from recent quark--meson models of nuclear
structure \cite{QMCIa,QMCIb,QMCII}, and those labeled ``QHD'' follow
from the present EFT for various parameter sets listed in
Table~\ref{tab:Gparams}.
It is obvious that the modern EFT approach improves the quality
of the fits by roughly \emph{two orders of magnitude\/} and establishes
a new standard of accuracy that must be attained for any modern
approaches to nuclear structure to be considered viable.

\begin{figure}
 \centering 
 \includegraphics*[width=3.0in,angle=-90.]{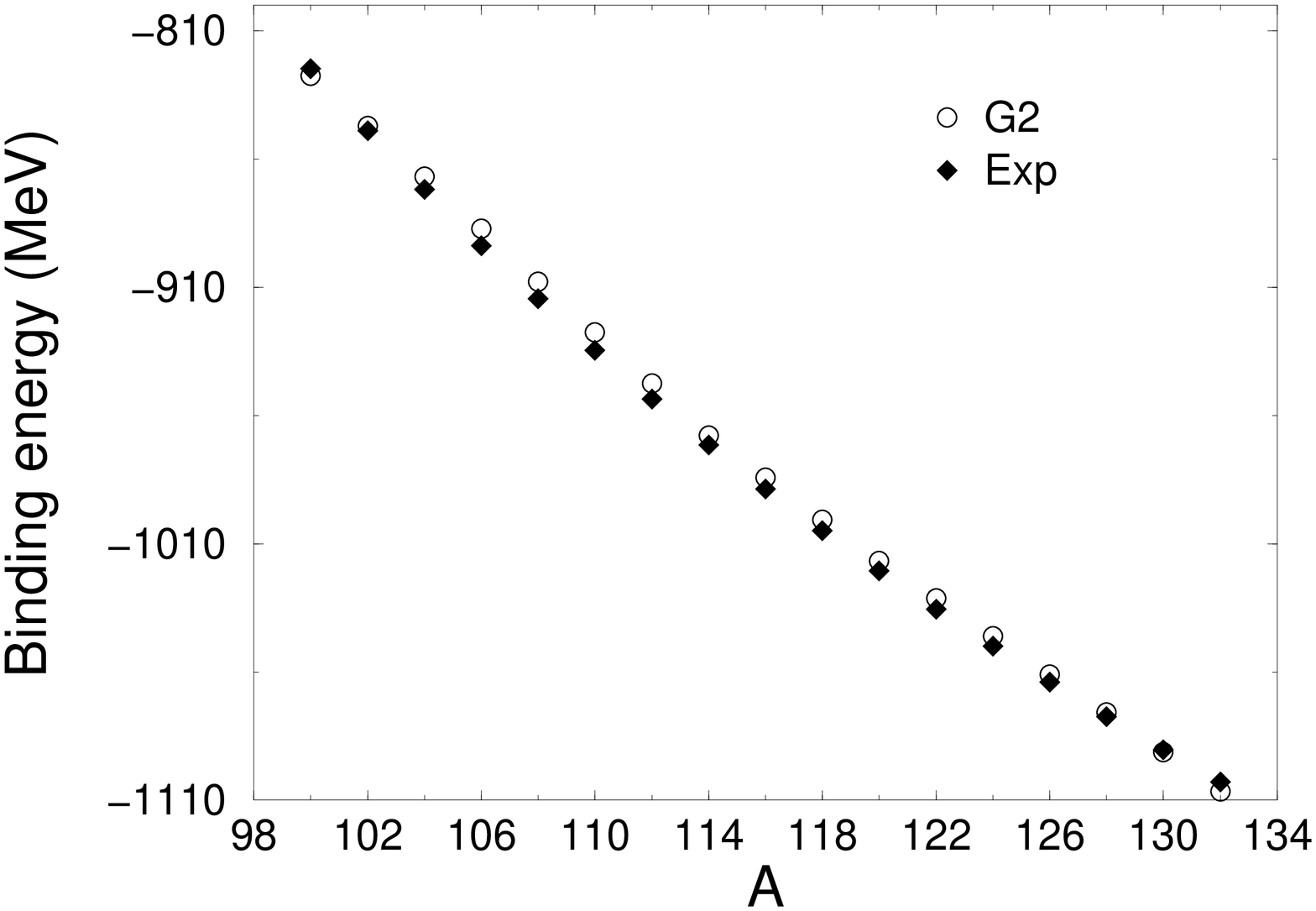}
 \caption{Binding energy of even-even 
          ${}^{\protect\mkern5mu A}_{50}$Sn isotopes
          calculated using the MFT of ${\cal L}_{\mathrm{QHD}}$
          with parameter set G2 from Table~\protect\ref{tab:Gparams}
          \cite{HUERTAS02}.
}
 \label{fig:snbe}
%
\vspace*{.5in}
%
 \centering 
 \includegraphics*[width=3.4in]{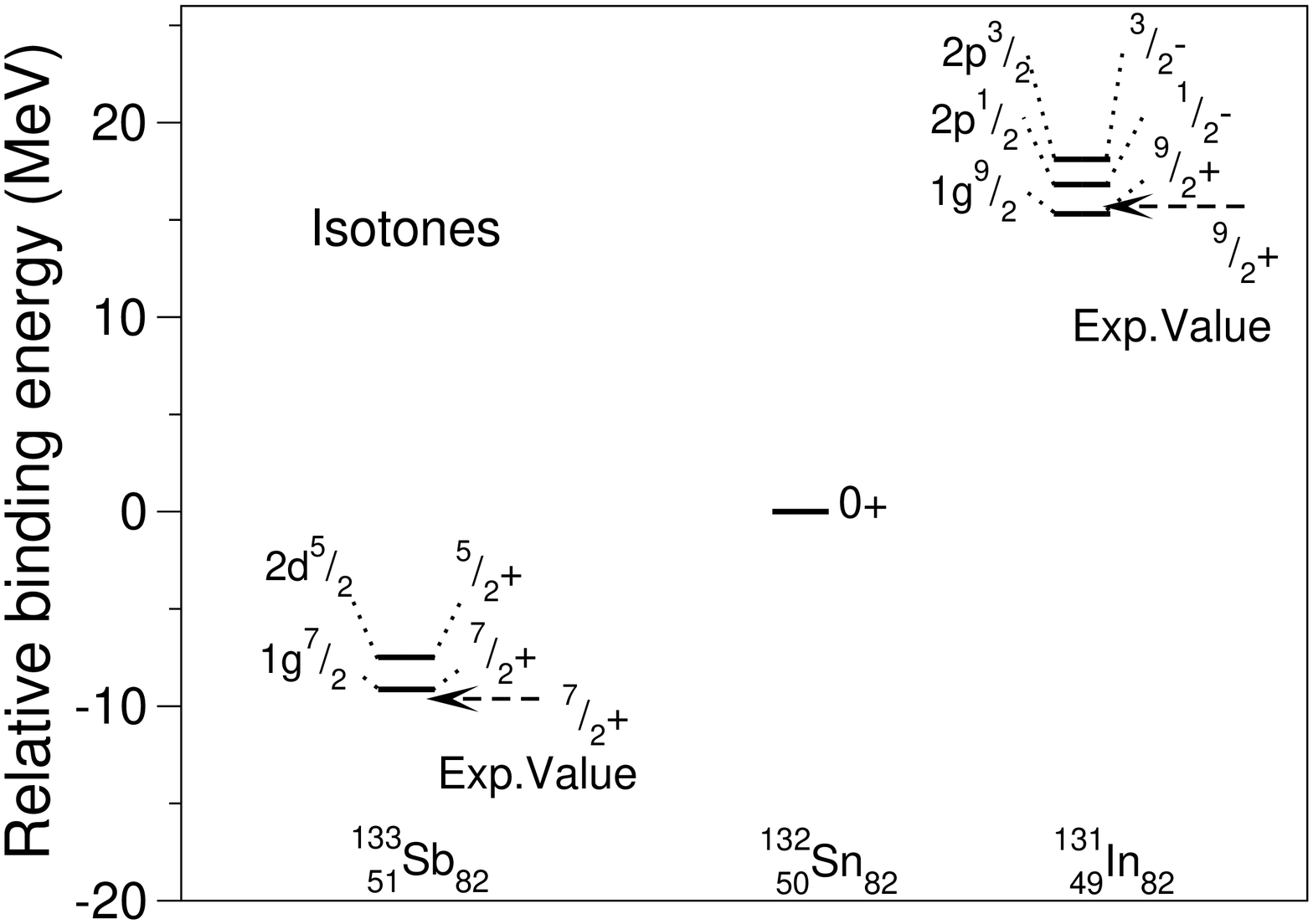}
 \caption{Calculated level spectrum of isotones of 
          ${}^{132}_{\protect\zz 50}$Sn${}_{82}$ 
          differing by one proton compared with empirical results
          (\emph{dashed lines with arrows\/}) \cite{HUERTAS02}.
}
 \label{fig:sn132isotones}
\end{figure}

Finally, to illustrate the power of the EFT approach to nuclear
many-body physics, we show two recent results of Huertas
\cite{HUERTAS02}.
The calculations use the MFT of ${\cal L}_{\mathrm{QHD}}$ with
parameters fitted to closed-shell nuclei along the ``valley of
stability'', namely, set G2 in Table~\ref{tab:Gparams}.
The resulting self-consistent, relativistic Kohn--Sham and meson
field equations are solved to calculate the properties of Sn
isotopes out to doubly-magic values of $N$ and $Z$ far from
stability.
These results are therefore true \emph{predictions} of the EFT,
since no adjustments were made to previously determined parameters.
Figure \ref{fig:snbe} shows the predicted binding energies of the
even-even isotopes from ${}^{100}_{\zz 50}$Sn to 
${}^{132}_{\zz 50}$Sn
compared with measured experimental results.
The theoretical values are accurate to better than 1\% throughout.
(Similar accuracy is obtained for parameter set G1; see Fig.~4 in
\cite{HUERTAS02}.)

Figure \ref{fig:sn132isotones} shows the predicted ground-state
energies, spins, and parities of the neighboring single-particle
and single-hole nuclei ${}^{133}_{\zz 51}$Sb and 
${}^{131}_{\zz 49}$In relative to ${}^{132}_{\zz 50}$Sn.  
The energy differences are just the chemical potentials, 
which should be accurately reproduced, according to the discussion 
of Kohn--Sham theory in Sect.~\ref{sec:DFT}.
The excellent agreement provides compelling evidence that QHD is 
indeed an EFT for low-energy QCD that can be used to describe nuclear
many-body physics.
This recent work has been extended to semi-magic nuclei with 
$N = 28, 50, 82, 126\:$ and $Z = 28, 50, 82\:$ in \cite{HUERTAS04}. 

How can we understand the excellent accuracy of the preceding MFT
results?
As discussed in Sect.~\ref{sec:DFT}, the exact energy functional
has kinetic-energy and Hartree parts (which are combined in a
relativistic formulation) plus an exchange-correlation functional,
which is generally a nonlocal, nonanalytic functional of the densities that
contains all the other many-body, relativistic, and short-range
effects \cite{FPW87,nmeos}.
The basic idea behind the relativistic MFT (RMFT) is to 
{\em approximate\/} the functional 
using an expansion in classical meson fields (or nucleon densities)
and their derivatives, based on the observation that the ratios of 
these quantities to the nucleon mass are small, 
at least up to moderate density.\footnote{Since the meson
fields are roughly proportional to the nuclear density, 
and since the spatial variations in nuclei are determined by the 
momentum distributions of the valence-nucleon wave functions, 
this organizational scheme is essentially an expansion in $\kf /M$, 
for $\kf$ corresponding to ordinary nuclear densities.
Here the nucleon mass $M \approx \varLambda$ is a generic large 
mass scale characterizing physics beyond the Goldstone bosons.}
The parameters introduced in the expansion are fitted to experiment,
and  if we have a systematic way to truncate the expansion, 
the framework is predictive.
Moreover, if the RMFT energy functional is sufficiently general, 
it will {\em automatically\/} incorporate effects beyond the Hartree 
approximation, such as those due to short-range physics
and many-body correlations.

But why should we expect an approximate, mean-field energy functional to 
work so well?
We observe that while the mean scalar and vector potentials $\Phi$ 
and $W$ are \emph{small\/} compared to the nucleon mass, 
they are large on nuclear energy scales \cite{BODMER91,FTS95}.
Moreover, as illustrated in 
Dirac--Brueckner--Hartree--Fock (DBHF) calculations 
\cite{RBBG,TERHAAR87,MACHLEIDT89},
the scalar and vector potentials (or self-energies) are nearly 
state independent and are almost equal to those obtained in the 
Hartree approximation.
Thus the Hartree contributions to the energy functional should 
dominate, and an expansion of the exchange-correlation functional 
in terms of mean fields should be reasonable.
This ``Hartree dominance'' also implies that it should be a good 
approximation to associate the single-particle Dirac eigenvalues 
with the empirical nuclear energy levels, at least for states near 
the Fermi surface \cite{DREIZLER90}.\footnote{%
One expects the KS spin-orbit splittings to be more accurate than the
absolute energy eigenvalues.}

We also observe that the nuclear properties of interest 
discussed in Sect.~\ref{sec:INTRO} include: 1)~nuclear 
shape properties, such as charge radii and charge densities, 
2)~nuclear binding-energy systematics, and 
3)~single-particle properties such as level spacings and orderings, 
which reflect spin-orbit splittings and shell structure.
Since the Kohn--Sham approach is formulated to reproduce
exactly the ground-state energy and density, and the Hartree 
contributions are expected to dominate the Dirac single-particle 
potentials, these observables are indeed the ones for which 
meaningful comparisons with experiment should be possible.

As discussed above, an RMFT energy functional of 
the form in (\ref{eq:vmdnm}), extended to include low-order 
derivatives of the meson fields, successfully reproduces these 
nuclear observables with parameters of natural size 
(see Table~\ref{tab:Gparams}).
This justifies a truncation of the energy functional at the first 
few powers of the fields and their derivatives, as is evident from
Fig.~\ref{fig:nmbe}.
Moreover, the full complement of parameters is underdetermined, 
so keeping only a subset does not preclude a realistic fit to nuclei.
Both the early RMFT calculations mentioned above and the newer 
calculations based on chiral EFT should be interpreted within the 
context of this Kohn--Sham approach to DFT.

\section{Analysis of Mean-Field-\kern-1pt Theory Parameters}
\label{sec:PARAMS}

Although we could use meson--nucleon EFT's as done historically, the
analysis is more transparent with point-coupling theories, which
contain only nucleon fields in a local Lagrangian.
Because of the freedom to perform field redefinitions, a general
point-coupling Lagrangian is equivalent to a general meson--nucleon
Lagrangian \cite{FST97r,RUSNAK97,FHT01}.

An energy functional of nucleon densities can be constructed by
starting with a general point-coupling effective Lagrangian, 
consistent with the symmetries of QCD, and by evaluating the
corresponding one-loop energy functional.
As discussed above, this approach approximates a general DFT
functional that incorporates many-body effects beyond the Hartree
level when the parameters are determined from finite-density
data.

To arrive at a suitable truncation scheme, we again rely on NDA and
naturalness.
The two relevant mass scales are $\fpi$ and $\varLambda$,
and for closed-shell nuclei, the energy functional is an expansion
in powers of the nucleon scalar, vector, isovector-vector, tensor,
and isovector-tensor densities, which are defined as
$\rhos \equiv \langle \psibar \psi \rangle$,
$\rhoB \equiv \langle \psi^{\dagger}\psi \rangle$,
$\rhothree \equiv {1\over 2} \langle \psi^{\dagger}
        \tau_3 \psi \rangle$,
$\stensor_i \equiv \langle \psibar \sigma^{0i} \psi \rangle$, and
$\stensor_{3i} \equiv {1\over 2} \langle \psibar \sigma^{0i} 
        \tau_3 \psi \rangle$, 
respectively, where $\psi$ is the nucleon field.

We can then define scaled densities and their derivatives as
[see (\ref{eq:NDAgen})]
\beq
\rhost \equiv \frac{\rhos}{f_\pi^2\varLambda} \ , \quad
{\wt \grad} \rhost \equiv \frac{\grad \rhos}{f_\pi^2\varLambda^2} \ ,
       \quad \mathrm{etc.}
   \label{eq:rhost}
\eeq
NDA also provides numerical estimates for the scaled densities that
will allow us to estimate terms in the energy functional.
For example, each additional power of $\rhos$ is accompanied by a
factor of $\fpi^2 \varLambda$.
The ratios of scalar and vector densities to this factor at nuclear
matter equilibrium density are between 1/4 and 1/7 \cite{FRIAR96},
which serves as an expansion parameter.
Similarly, one can anticipate good convergence
for gradients of the densities, since the relevant scale for
derivatives in finite nuclei is the nuclear surface thickness $\sigma$, 
and so the dimensionless expansion parameter 
is $1/\varLambda\sigma \le 1/5$.
The expansion is useful because the coefficients have been shown
empirically to be natural, that is, of order unity 
\cite{SW97r,RUSNAK97}.

The energy functional is dominated by the isoscalar terms, but we also
include the isovector and tensor terms for completeness.
To nominal order $\nu = 4$ (all densities and gradients count as $\nu = 1$,
except the three-vector tensor densities 
$\mbox{\boldmath{$\tensorN$}}$, which have 
$\nu \approx 2$; see also footnote \ref{fn:alphas}
on p.~\pageref{fn:alphas}), 
we obtain
%

\beqa
E &=&
  \sum_{\alpha}^{\rm occ} \int\! {\D}^3x\; \psibar_\alpha(
     -i \beta\, \mbox{\boldmath $\alpha\,\cdot\,$}\grad + M)\psi_\alpha 
 \nonumber\\[-2pt]
 & &  
       {} + f_\pi^2 \varLambda^2 \!\!\int\!{\D}^3x\,\bigg\{
         \wt\kappa_2\rhost^{\smk 2}
        -\wt\kappa_{\rm d}  ({\wt\grad}\rhost)^2  
   +\wt\kappa_3\rhost^{\smk 3}
   +\wt\kappa_4\rhost^{\smk 4}
   +\wt\eta_1 \rhoBt^{\smk 2}\rhost
   +\wt\eta_2 \rhoBt^{\smk 2}\rhost^{\smk 2}
   \nonumber\\[4pt]
 & &  \qquad\qquad\qquad\ {}+\wt\zeta_2\rhoBt^{\smk 2}
    -\wt\zeta_{\rm d} ({\wt\grad}\rhoBt)^2 
    +\wt\zeta_4 \rhoBt^{\smk 4}
    -\wt\alpha_1 \rhost  ({\wt\grad}\rhost)^2
    -\wt\alpha_2 \rhost  ({\wt\grad}\rhoBt)^2
     \nonumber\\[6pt]
 & &  \qquad\qquad\qquad\ {}+ \wt\xi_2\rhothreet^{\smk 2}
    -{\wt\xi}_{\rm d}  ({\wt\grad}\rhothreet)^2 
    +\wt\eta_\rho \rhothreet^{\smk 2}\rhost
    +\wt f_{\rm v}{\wt\grad}\rhoBt 
    \mbox{\boldmath$\,\cdot\,\tensorN$}
    +\wt f_\rho
	{\wt\grad}\rhothreet \mbox{\boldmath$\,\cdot\,\isovectorTensorN$}
	\nonumber\\[4pt]
 & & \qquad\qquad\qquad\ 
     {} + \ \mbox{electromagnetic\ and\ higher-order\ terms}
        \bigg\} \ , \qquad
 \label{eq:functfull} 
\eeqa
where the notation of \cite{RUSNAK97} is used.
The sum runs over occupied nucleon states.

\begin{figure}
  \begin{center}
\vspace*{0.2in}
  \includegraphics*[width=3.5in]{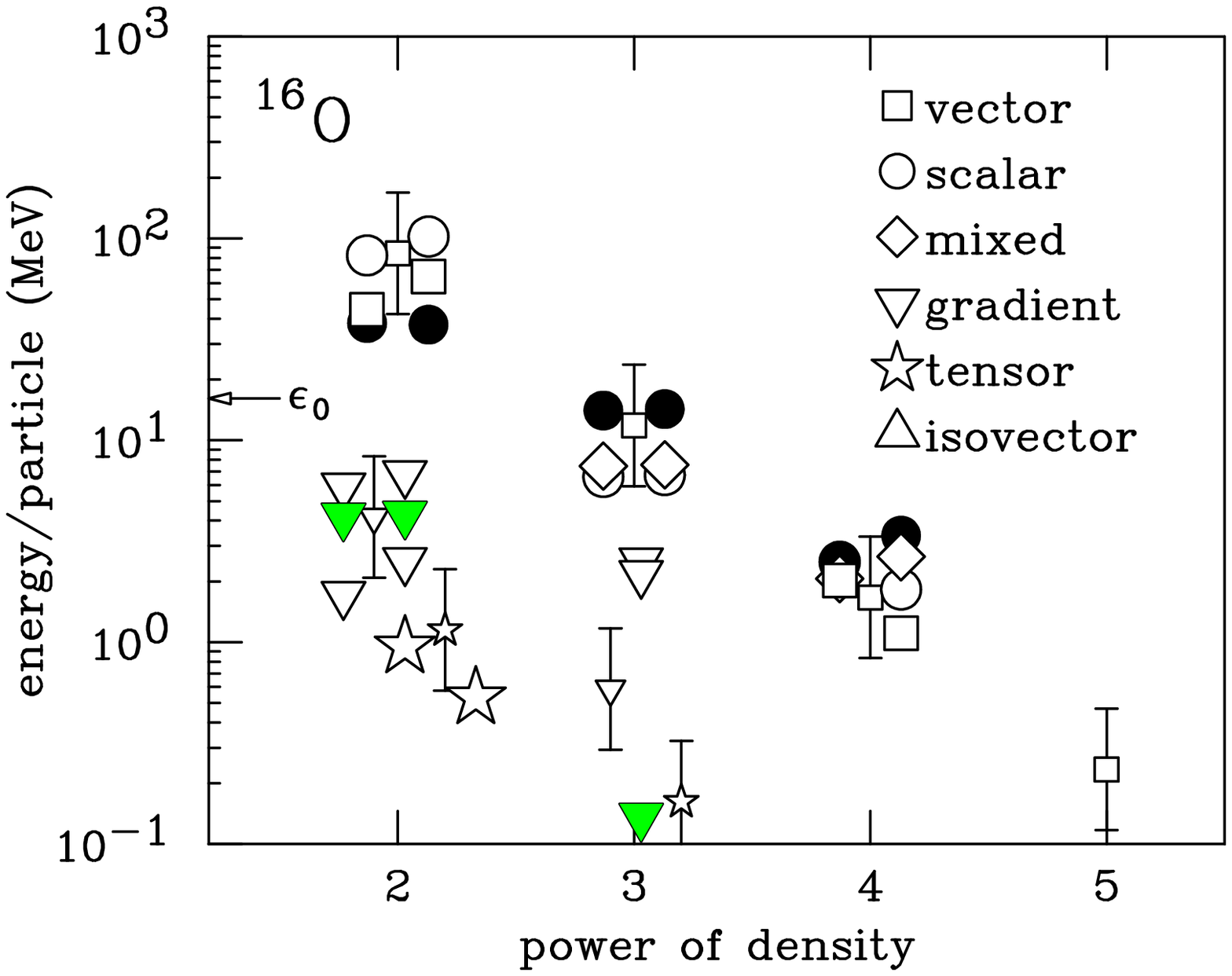}
  \end{center}
\vspace*{0.1in}
  \begin{center}
  \includegraphics*[width=3.5in]{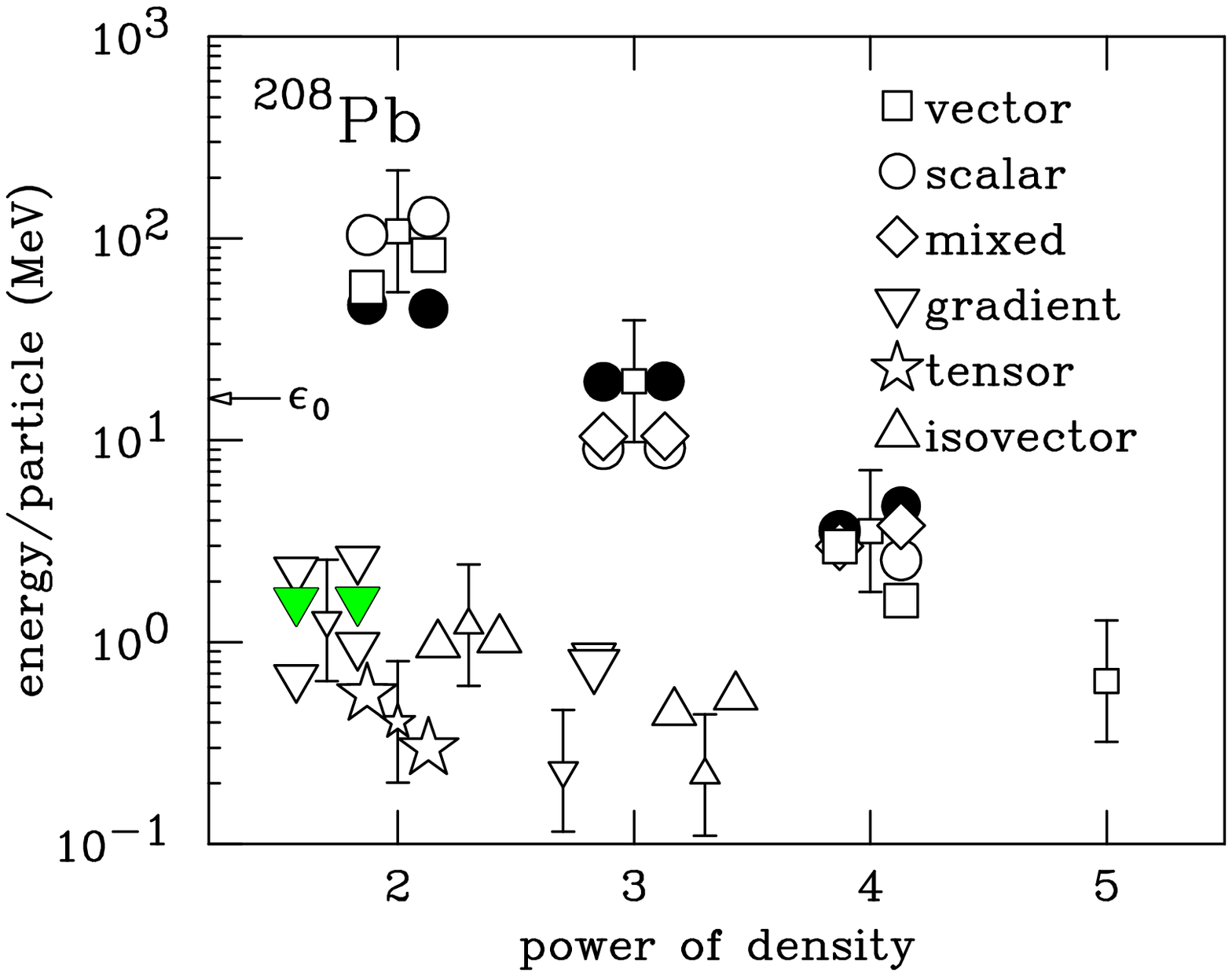}
  \end{center}
  \caption{Contributions to the energy/particle in ${}^{16}$O and
           ${}^{208}$Pb determined by logarithmic derivatives with respect 
           to the model parameters (see text) for two RMF point-coupling
           models \protect\cite{RUSNAK97}.
           Absolute values are shown.
           The filled symbols are net values.
           The small symbols indicate estimates based on NDA \cite{FSp00r},
           with the error bars corresponding to natural coefficients from
           1/2 to 2.
}
  \label{fig:be_pc_o16}   
\end{figure}

In Fig.~\ref{fig:be_pc_o16}, the small symbols show the NDA estimates
(with associated error bars) for the various energy contributions in
${}^{16}$O and $^{208}$Pb.
The magnitudes of energy contributions in (\ref{eq:functfull}) 
from two representative RMF point-coupling models (i.e., two different
parameter sets) are shown as larger unfilled symbols (one model on each 
side of the error bars).
These models provide very accurate predictions of bulk 
nuclear properties \cite{RUSNAK97}.
The energy contributions are determined for each nucleus by making multiple
runs while varying each parameter slightly around its optimized
value, which enables us to deduce the logarithmic derivative
with respect to each parameter.
The filled symbols denote the sum of the values for each power of 
the density.
The binding energy/nucleon in equilibrium nuclear matter is 
denoted by $\epsilon_0$.

The two representative point-coupling models validate the isoscalar 
estimates (small open squares), and  the resulting hierarchy of isoscalar 
contributions is quite clear.
How far down in the hierarchy can we reliably
determine contributions and their associated parameters?
In Fig.~\ref{fig:chisq}, the impact of different truncations of RMF 
meson--nucleon models is shown by plotting the figure of merit 
against the maximum power of fields.
We have also performed this test with point-coupling models,
with similar conclusions.
The ``full'' models (which include all nonredundant terms at a given order) 
show that one needs to go to the fourth power of the fields to get the best 
fits, but going further yields no improvement.
Analogous behavior is found for RMF point-coupling models with
powers of densities replacing powers of fields \cite{RUSNAK97}.
Thus contributions to the energy/particle at the level of
roughly 1\,MeV are at the limit of resolution.
Fifth-order isoscalar contributions to the energy/particle,
which are predicted to be less 
than 1\,MeV, are simply not determined by the optimization.

\begin{figure}
  \centering
  \includegraphics*[width=3.in]{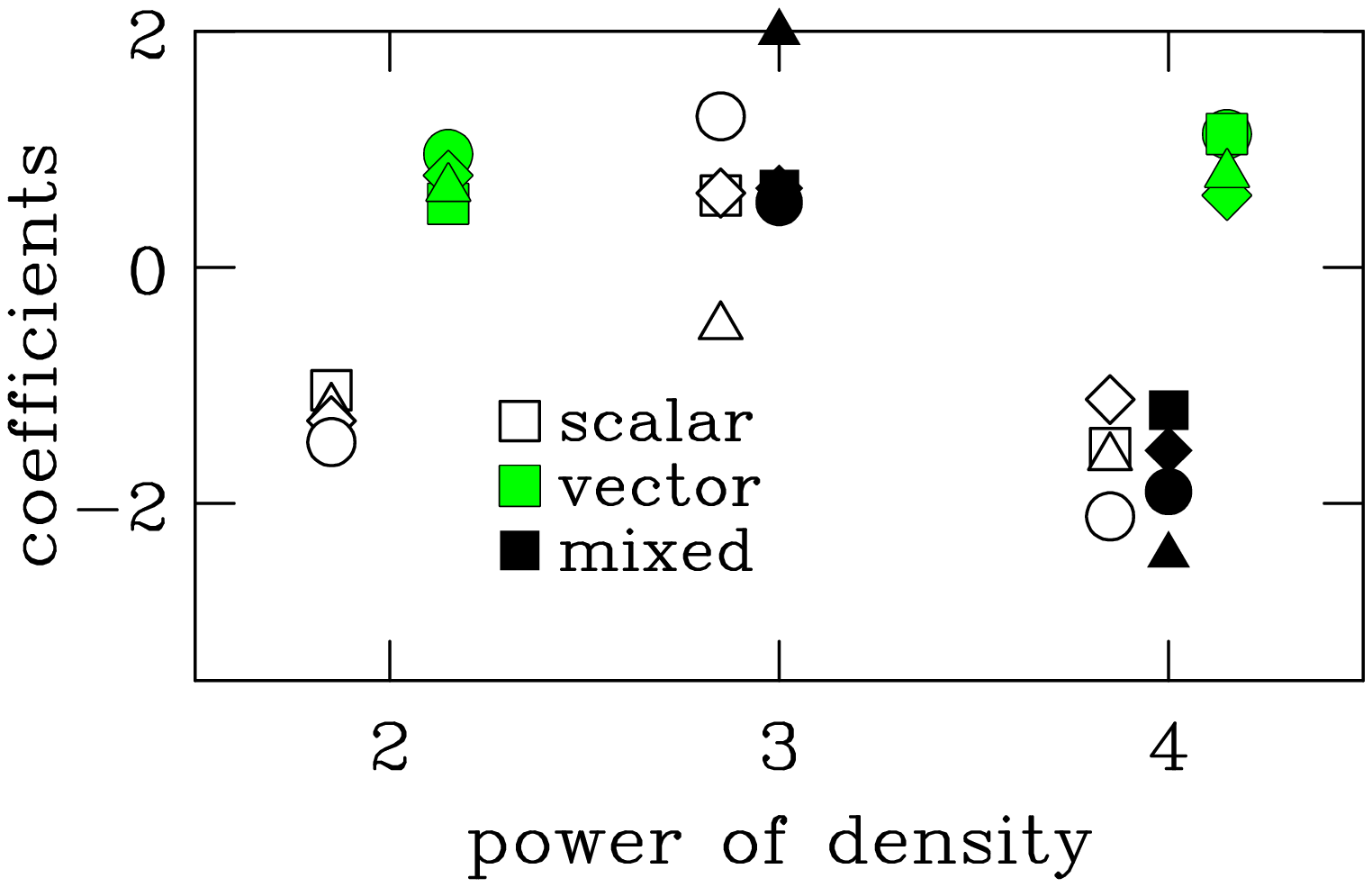}
\caption{Coefficients for four accurately fit RMF point-coupling
         models from \protect\cite{RUSNAK97}.  
         Each model is represented by a different shape,
         and the shading shows the type of term 
         (scalar, vector, or mixed).}
    \label{fig:coeffspc}
\end{figure}

The variation in coefficient values in (\ref{eq:functfull}) provides 
a measure of how well the parameters are actually determined by the data.
Figure~\ref{fig:coeffspc} shows the seven coefficients of isoscalar 
non-gradient terms from four point-coupling models.
Note that all coefficients are natural, i.e., of order unity.
However, the spread in coefficient values is significant and does not 
correspond well to the power-counting order.
We conclude that \emph{different linear combinations\/} of the coefficients
must be considered to draw reliable conclusions about how many
are determined by the data.

%
\begin{table}
\centering
\renewcommand{\baselinestretch}{1.0}
\caption{Improved coefficients for point-coupling RMF models
         (Table~VI of \protect\cite{RUSNAK97}).}
\vspace{.1in}
\begin{tabular}{cccc}
\hline\hline\\[-4pt]
            & linear      & density & deduced   \\
coefficient & combination & scaling & value  \\
\hline\\[-5pt]
  $\wt\Omega_1$ & $\wt\kappa_2+\wt\zeta_2$ & $\rhoplus$ & $-0.51\pm 0.01$\\
 $\wt\Omega_3$ & $\wt\kappa_3+\wt\eta_1$ & $\rhoplus^2$ & $+1.3\pm 0.1$ \\
 $\wt\Omega_2$ & $\wt\kappa_2-\wt\zeta_2$ & $\rhominus$ & $-2.0\pm 0.4$ \\
 $\wt\Omega_5$ & $\wt\kappa_4+\wt\zeta_4+\wt\eta_2$ & $\rhoplus^3$ 
     & $-2.4\pm 0.7$ \\
 $\wt\Omega_4$ & $\wt\kappa_3-\wt\eta_1/3$ & $\rhoplus\rhominus$ 
     & $+0.2\pm 1.0$ \\
 $\wt\Omega_6$ & $\wt\kappa_4-\wt\zeta_4$ & $\rhoplus^2\rhominus$ 
     & $-2.6\pm 0.8$ \\
 $\wt\Omega_7$ & $\quad\wt\kappa_4+\wt\zeta_4-2\kern1pt \wt\eta_2 /3 \quad$ &
         $\quad\rhoplus\rhominus^2 \quad$  &  
 \\[4pt] 
 \hline\hline
\end{tabular}
\label{tab:improved}
\end{table}

\begin{figure}
  \centering
  \includegraphics*[width=3.3in]{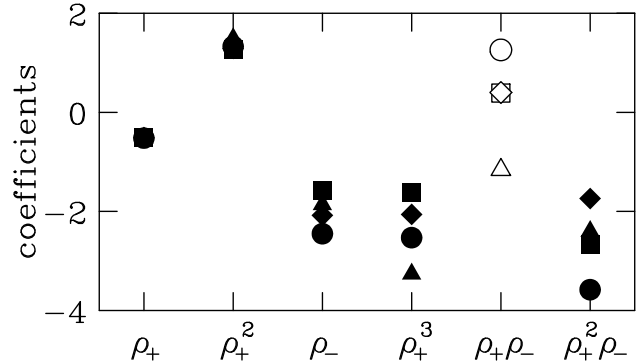}
\caption{Improved coefficients for the same four models as
         in Fig.~\ref{fig:coeffspc}.
         The ``order'' is determined by counting powers of
         $\rhoplus$ and $\rhominus$.}
   \label{fig:optimalpc}
\end{figure}

Can we find a more systematic power counting scheme?
The similar size of the scalar density $\rhos$ and
the vector density $\rhoB$ suggests that we
count instead powers of $\rhoplus \equiv (\rhos + \rhoB)/2$ and
$\rhominus \equiv (\rhos - \rhoB)/2$.
The corresponding ``improved'' coefficients are listed in
Table~\ref{tab:improved} [see (\ref{eq:functfull})].
The spread in these coefficients for four RMF point-coupling models from 
\cite{RUSNAK97} is shown in Fig.~\ref{fig:optimalpc}.
The terms are organized according to the powers of $\rhoplus$
and $\rhominus$, with $\rhominus$ scaling as $\rho_+^{8/3}$.

The leading orders are very well determined, with a systematic increase 
in uncertainty.
Even the sign is undetermined for the parameter $\wt\Omega_4$,
which is shown with unfilled symbols, but the next parameter ($\wt\Omega_6$) 
appears to be reasonably well determined.
Higher-order terms are not determined by the optimizations.
Deduced values and uncertainties based on this sample of models
are given in Table~\ref{tab:improved}.
We see that of the seven isoscalar non-gradient parameters in
(\ref{eq:functfull}), four linear combinations are clearly
determined by bulk nuclear observables, with probably a fifth combination 
as well.

The isovector terms appear only on the graphs for $^{208}$Pb.
The factor $(N-Z)/2B$, which is only 10\% even for Pb, severely limits
the sensitivity to isovector terms (especially since the factor must 
appear with even powers).
The magnitude of the leading isovector term 
($\propto \wt\xi_2 \rhothreet^{\smk 2} \,$)
is comparable to the fourth-order isoscalar
term, which is at the limit of what can be determined reliably 
from fitting the binding energy.
We conclude that only one isovector parameter is determined by the bulk 
observables.
For a further discussion of the isovector terms, see \cite{FURNSTAHL02}.

The energy estimates for isoscalar, tensor terms imply that only one
parameter, at best, can be determined.
Higher-spin terms, which will require more gradients and have smaller
average densities, are not at all constrained.
The tensor terms are interesting because a fraction of the spin-orbit force 
can be generated by including an isoscalar, tensor coupling of the vector 
field to the nucleon \cite{FRS98r}.
Nevertheless, the spin-orbit potential arises predominantly from 
the large scalar and vector fields; attributing more than one-third
of the potential to the tensor coupling produces unrealistic
surface systematics \cite{HUA99}. 
 
Finally, gradient terms follow the same pattern in the energy:
the leading term is barely above the limit of resolution.  
In fact, there are two isoscalar gradient terms at leading order 
(scalar and vector), but only their sum is well determined.  
The sum of the subleading-order contributions almost vanishes.
We conclude that only one gradient parameter can be 
determined \cite{FURNSTAHL02}.

The handful of parameters that are well determined by the usual bulk nuclear
observables (binding energies, charge density distributions, and spin-orbit 
splittings in doubly magic nuclei) can be associated with an equal number 
of nuclear properties and general features of RMFT's. 
In particular,  
\begin{enumerate}
   \item Two isoscalar non-gradient parameters are very well determined.
   These correspond to the highly constrained values for
   the equilibrium density ($\kf \approx 1.30\pm 0.01\,\mbox{fm}^{-1}$) 
   and binding energy ($16.0\pm 0.1\,\mbox{MeV}$)
   of nuclear matter.
   \item An additional isoscalar constraint is that 
   $\Mstar \approx 0.61\pm 0.03$, if the isoscalar, tensor term is set 
   to zero.
   This range ensures an accurate reproduction of spin-orbit splittings
   in finite nuclei.
   Small increases in $\Mstar$ without changing the splittings
   can be accomplished by including an isoscalar, tensor term; an analysis 
   using a simple local-density approximation is discussed
   in \cite{FRS98r}.
   \item A fourth isoscalar constraint comes from the
   nuclear matter compressibility.  
   The constraint is much weaker, in the range of 
   $K \approx 250 \pm 50\,$MeV.
   \item The possibility of a fifth isoscalar constraint has been
   considered by Gmuca \cite{GMUCA92}, who argued that 
   separate scalar and vector fourth-order terms were necessary to  
   tune the density dependence of the scalar and vector parts of the 
   baryon self-energy.
   This would correspond to constraining $\wt\Omega_6$ from
   Table~\ref{tab:improved}.  
   Moreover, some form of isoscalar nonlinear vector interaction is 
   needed to soften the high-density equation of state to be consistent 
   with observed neutron star masses \cite{nmeos}.
   \item Since only one isoscalar gradient parameter is determined,
   it is not useful to allow the scalar and vector masses 
   (or their equivalents in a point-coupling theory) to
   vary independently.  
   Thus it is convenient to fix the vector mass at a natural size, 
   such as the experimental mass for the $\omega$.
   A scalar mass of  $500\pm 20\,$MeV is then required.
   \item The one isovector parameter can be fixed by
   the surface-corrected volume symmetry energy \cite{SEEGER75}, 
   which falls in the range $34\pm 4\,$MeV \cite{HS81}.  
   Since no isovector gradient is determined, setting the isovector,
   vector meson mass to the experimental $\rho$ meson mass is adequate.  
\end{enumerate}

To ensure a reasonable (if not optimal) description of finite nuclei,
it is sufficient to reproduce the nuclear matter properties given above,
but note that \emph{all\/} calculated properties must be accurate, and the 
resulting parameters must be natural.%
\footnote{A further caution is that there are many correlations among
these properties, so that the allowed ranges should not be considered to
be independent.}
One cannot justify the underlying physics of a model
if it reproduces only a subset of the nuclear calibration data.

\finalnewpage 
\section{Weak Nuclear Currents}
\label{sec:WNC}

A desirable theory of nuclear currents
should satisfy the following three criteria:
\begin{itemize}
\item
It should use the same degrees of freedom to describe the currents
and the strong-interaction dynamics.
\item
It should respect the same internal symmetries, both discrete and 
continuous, as the underlying theory of QCD.
\item
Its parameters can be calibrated using strong-interaction phenomena, 
like $\pi$N scattering and the properties of finite nuclei.
This is especially important in EFT's, as they contain all (non-redundant) 
interaction terms that are consistent with the underlying symmetries 
\cite{WEINBERGone,SW97r}.
\end{itemize}
\emph{The \emph{QHD} framework described so far embodies these three
desirable features}.

\begin{figure}
 \centering 
 \includegraphics*[width=3.37in]{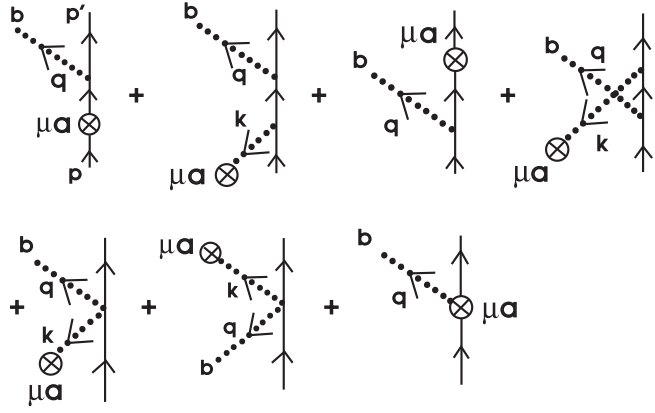}
   \caption{The amplitude for axial-current pion production 
            on a single nucleon [Eq.~(\protect\ref{eq:piprod-2})].
            Note that momentum $k^{\mu}$ is \emph{extracted\/} by the 
            external source (which is denoted by $\otimes$),
            so that $p^{\mu} = {p'}^{\mu} + k^{\mu} + q^{\mu}$.}
\label{fig:pi-pi-n}
\end{figure}

The weak currents arise from Noether's theorem applied directly to 
${\cal L}_{\mathrm{QHD}}$ and contain the pion field to all orders.
The leading-order (in $\nu$) vector and axial-vector currents are given 
by ($a$ is the isospin index)
\beqa
V^{a \mu} & = &
  - i\,{ {f_{\pi}^2} \over 4} \, {\rm Tr} 
\left\{ {\tau}^a \left( U {\partial}^{\mu} U^{\dag} + U^{\dag}
{\partial}^{\mu} U \right) \right\} 
+ { {1} \over {4} }\, \Nbar {\gamma}^{\mu} 
\left[ {\xi} {\tau}^a  {\xi}^{\dag} + {\xi}^{\dag} {\tau}^a 
 {\xi} \right] N 
\nonumber \\[3pt]
 & & \quad
 {} + { {1} \over {4} }\, g_A \Nbar {\gamma}^{\mu} {\gamma}_5
\left[ {\xi} {\tau}^a  {\xi}^{\dag} - {\xi}^{\dag} {\tau}^a 
 {\xi} \right] N \ , \\[3pt]
%
A^{a \mu} & = &
  - i\,{ {f_{\pi}^2} \over 4}\, {\rm Tr} 
\left\{ {\tau}^a \left( U {\partial}^{\mu} U^{\dag} - U^{\dag}
{\partial}^{\mu} U \right) \right\} 
- { {1} \over {4} } \,\Nbar {\gamma}^{\mu} 
\left[ {\xi} {\tau}^a  {\xi}^{\dag} - {\xi}^{\dag} {\tau}^a 
 {\xi} \right] N
\nonumber \\[3pt] 
 & & \quad 
 {} - { {1} \over {4} }\, g_A \Nbar {\gamma}^{\mu} {\gamma}_5
\left[ {\xi} {\tau}^a {\xi}^{\dag} + {\xi}^{\dag} {\tau}^a 
 {\xi} \right] N \ .
%
\eeqa
%
As discussed in \cite{AXC}, the Feynman rules for the lowest-order
$\pi$N vertices can be determined from (\ref{eq:eft-lagrangian}),
and in the presence of an external axial-vector source, one can
compute the one- and two-nucleon amplitudes for the axial-vector 
current.
The amplitude for axial-current pion production on a
single nucleon is represented by the diagrams of
Fig.~\ref{fig:pi-pi-n} and can be written as
\begin{eqnarray}
 2 f_{\pi} \: M^{ab\, \mu} ( \pi ) & = & 
 \bar{u}(p') \left\{ g^2_A \left[ \not \! q \, {\gamma}_5 \, {1 \over
 {(\not \! p^{\mkern2mu \prime} \, 
 + \not \!  q)-M}} \, {\gamma}_5 \left( {\gamma}^{\mu} -
 \not \! k { {{k}^{\mu}} \over {k^2 - m_{\pi}^2} } \right) {
 { {\tau}^b {\tau}^a } \over 2} \right. \right. \nonumber \\[6pt] 
&  & \qquad\qquad +
 \left. \left( {\gamma}^{\mu} - \not \! {k} { {k^{\mu}} \over {k^2 -
 m_{\pi}^2} } \right) {\gamma}_5 {1 \over {(\not \! p \; -
\not \! q) - M}} \,{\gamma}_5 \!\not \! q \,
 { { {\tau}^a {\tau}^b } \over 2} \right] \nonumber \\[6pt] 
&  & \qquad\qquad
 + \left. i {\epsilon}^{abc} \, { {{\tau}^c} \over 2} \left[ (\not
 \! k\, - \!\not \! q) { {k^{\mu}} \over {k^2 - m_{\pi}^2}} - 2
 {\gamma}^{\mu} \right] \right\} u(p) \ ,
   \label{eq:piprod-2}
\end{eqnarray}
where $q^{\mu}$ is the outgoing four-momentum of the emitted pion,
and $p^{\mu}$ and ${p'}^{\mu}$ are the initial and final
nucleon four-momenta, respectively.

\begin{figure}
 \centering 
 \includegraphics*[width=2.8in]{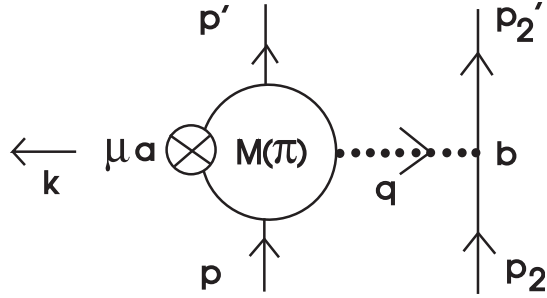}

  \caption{Two-nucleon contributions to the axial current.
        Here the vertex $M(\pi )$ 
        [Eq.~(\protect\ref{eq:piprod-2})]
        represents all possible
        ways that the external source can couple to the
        left-hand nucleon line (to the order we are working),
        as illustrated in Fig.~\protect\ref{fig:pi-pi-n}.
        There are also contributions from the right-hand nucleon
        line, as well as exchange terms.}
\label{fig:pi-a-2n0}
\end{figure}

The two-nucleon amplitude for the axial current, to this order in $\nu$,
is given in Fig.~\ref{fig:pi-a-2n0}.
It is easy to verify that this amplitude satisfies PCAC 
(for on-shell nucleons).

\begin{figure}
 \centering 
 \includegraphics*[width=3.0in]{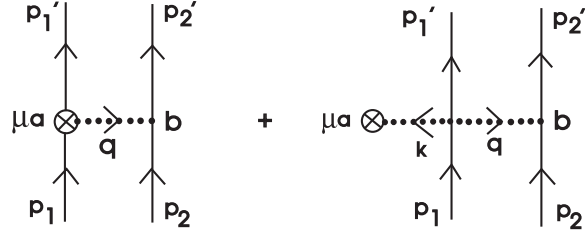}
  \caption{Two-nucleon, axial-current amplitudes originating
           from the $\nu = 3$ pion--nucleon terms in the 
           effective Lagrangian.
           There is one pair of diagrams proportional to
           $\kappa_\pi$ and another pair proportional to $\beta_\pi$.}
\label{fig:pi-a-2n}
\end{figure}

\begin{figure}
 \centering
\vspace{0.2in} 
 \includegraphics*[width=3.0in]{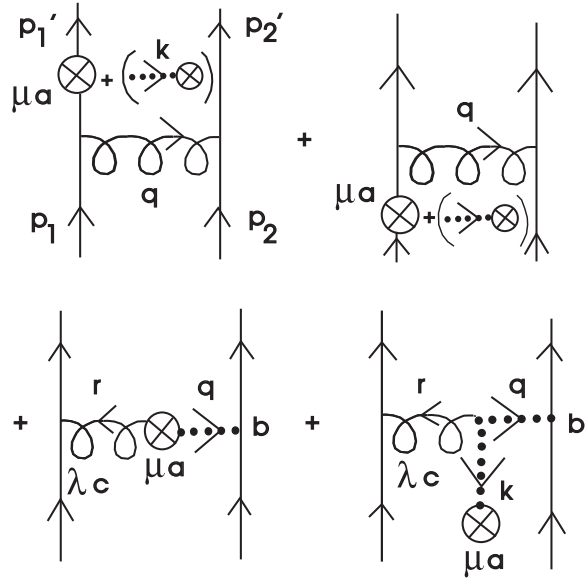}
   \caption{Two-nucleon, axial-current amplitudes containing
            rho meson exchange.
            The notation in the first two diagrams implies that
            the external source can couple to the nucleon line
            in two ways, as in Fig.~\protect\ref{fig:pi-a-2n}.}
\label{fig:rho-a-2n}
\end{figure}

One can also determine the vector and axial-vector
currents arising from the $\nu = 3$
and $\nu = 4$ terms in the QHD Lagrangian (\ref{eq:eft-lagrangian}).
These additional contributions take the form
\beqa
\delta V^{a \mu} & = &
  {- {i {\kappa}_{\pi}} \over {2 M} } \, \Nbar
{\sigma}^{\mu \nu} \left[ {\xi}^{\dag} {\tau}^a \xi - \xi {\tau}^a
{\xi}^{\dag}, \: a_{\nu} \right] N - { {2 {\beta}_{\pi}} \over {M} }
\, \Nbar N \, {\rm Tr} \left[ ( {\xi}^{\dag} {\tau}^a \xi - \xi
{\tau}^a {\xi}^{\dag} ) \, a^{\mu} \right] \nonumber \\[5pt]
 & & \quad
 {} - {\epsilon}^{abc} {\rho}^b_{\nu}
 \left\{ { {f_{\rho} g_{\rho}} \over {4M} } \, \Nbar {\sigma}^{\mu
 \nu} {\tau}^c N + \left( {\partial}^{\mu} {\rho}^{\nu }_c 
 - {\partial}^{\nu} {\rho}^{\mu }_c \right)
 + O({\pi}^2 ) \right\} 
\eeqa
and
\beqa
\delta A^{a \mu} & = &
  {- {i {\kappa}_{\pi}} \over {2 M} } \, \Nbar {\sigma}^{\mu \nu}
\left[ {\xi}^{\dag} {\tau}^a \xi + \xi {\tau}^a {\xi}^{\dag}, 
\: a_{\nu} \right] N
- { {2 {\beta}_{\pi}} \over {M} } \, \Nbar N \, {\rm Tr}
\left[ ( {\xi}^{\dag} {\tau}^a \xi 
+ \xi {\tau}^a {\xi}^{\dag} ) \, a^{\mu} \right] \nonumber \\[5pt]
 & & \quad
 {} + \left\{
   g_{\rho \pi\pi} \, { {2 f_{\pi}} \over {m_{\rho}^2} } \, 
{\epsilon}^{abc} \,
{\partial}_{\nu} {\pi}^b \left( {\partial}^{\mu}
{\rho}^{\nu}_c - {\partial}^{\nu}
{\rho}^{\mu}_c \right) + O(\rho^2 \pi , \, \rho \pi^3 ) \right\} \ .
\eeqa
The terms proportional to $\kappa_\pi$ and $\beta_\pi$ produce the
two-nucleon amplitudes shown in Fig.~\ref{fig:pi-a-2n}.
The corresponding amplitudes involving rho meson exchange are
illustrated in Fig.~\ref{fig:rho-a-2n}, and there are also amplitudes
with isoscalar scalar and vector meson exchange that resemble the
first two diagrams in this figure.
Analytical expressions for all of these two-nucleon amplitudes are 
given in \cite{AXC}.

The scattering amplitudes constructed from the currents 
listed above satisfy
CVC, PCAC (when $m_{\pi} \not= 0$), and the Goldberger--Treiman
relation (with $\gA \not= 1$) \emph{automatically}.
Moreover, the chiral charges $Q^a$ and $Q_5^a$ derived from these currents
satisfy the familiar chiral charge algebra \emph{to all orders in the
pion field} \cite{AXC}.
In \cite{HUERTASwk}, these currents are used to study beta decay in 
$^{131,133}$Sn.

\section{Summary}
\label{sec:SUMM}

In this work, I discussed recent progress in Lorentz-covariant quantum 
field theories of the nuclear many-body problem, often called quantum 
hadrodynamics (QHD).
QHD is a local, nonrenormalizable, effective Lagrangian field theory with 
baryons and mesons as the generalized coordinates (fields).
An effective Lagrangian consists of known long-range interactions
constrained by symmetries and a complete, non-redundant set of short-range
interactions.
By simply looking at the spectra of massive nuclei, it is obvious that 
\emph{some\/} 
relativistic effects must be important in nuclei; thus, it is most 
convenient to use a Lorentz-covariant theory.

The effective field theory studied here contains nucleons,
pions, isoscalar scalar ($\sigma$) and vector ($\omega$) fields, and
isovector vector ($\rho$) fields.
The heavy mesons are introduced as collective, effective degrees of freedom
to simplify the description of the medium- and short-range
nucleon--nucleon interaction and to conveniently parametrize ground-state
expectation values of nucleon bilinears, which are important for
the description of bulk nuclear properties.
The QHD theory exhibits a nonlinear realization of 
spontaneously broken $\mathrm{SU(2)}_L \times \mathrm{SU(2)}_R$ 
chiral symmetry and has three desirable features:
it uses the same degrees of freedom to describe the nuclear currents
and the strong-interaction dynamics,
it respects the symmetries of the underlying theory of QCD, 
and its parameters can be calibrated using strong-interaction phenomena.
Moreover, the electromagnetic structure of the nucleon can be included
straightforwardly in a derivative expansion of the fields.

Although the QHD Lagrangian in principle contains an infinite number of
terms, naive dimensional analysis and naturalness
allow one to identify suitable expansion parameters and to estimate the
sizes of various terms in the Lagrangian.
Thus, for any desired accuracy, the Lagrangian can be truncated to a
finite number of terms.
In particular, for normal nuclear systems, it is possible to expand the 
QHD effective Lagrangian systematically in powers of the meson fields 
(and their derivatives) and to truncate the expansion reliably after 
the first few orders.

Using density functional theory, I showed that the mean-field approximation
produces an energy functional whose parameters can be determined by fitting
bulk and single-particle properties of nuclei.
The framework of Kohn--Sham theory allows the ground state to be constructed
from (quasi)particle orbitals with unit occupation number.
Since the mean-field energy functional is a good approximation to the
exact energy functional over the relevant range of density, it is possible
to reproduce nuclear densities, binding energies, and single-particle 
spectra near the Fermi surface very accurately.
Because the parameters are fitted to nuclear properties, the energy
functional implicitly contains effects that go beyond a simple Hartree
approximation, such as short-range physics, hadron substructure,
and many-body correlations.

The numerical parameters of QHD were studied using an effective, 
point-coupling Lagrangian that contains nucleon fields only.
Because of the freedom to redefine the fields (or coordinates), a general
point-coupling theory is equivalent to a general baryon--meson theory.
By examining the contributions to the energy/nucleon in doubly magic
nuclei, it was found that only a small number of parameters (roughly 
seven) can be calibrated by the nuclear data input.
New ways to calibrate additional parameters will play an important role
in the construction of the next generation of QHD Lagrangians.
Finally, the weak vector and axial-vector currents, and the corresponding
two-nucleon, axial-current amplitudes,
were discussed in the QHD framework.

Nuclear physics is the study of strongly interacting
hadronic matter, and the only consistent theoretical framework we have
for describing such a relativistic, interacting, quantum-mechanical,
many-body system is relativistic quantum field theory based on a local
Lagrangian density.
Although QCD of quarks and gluons provides the basic underlying theory,
Lagrangians comprised of hadronic degrees of freedom (QHD)
provide the most efficient description of the physics in the 
strong-coupling domain.
In the modern effective field theory perspective of QCD, one incorporates
only the underlying symmetries of QCD into the QHD Lagrangian.
By interpreting the mean-field approximation in the context of density
functional theory, one can understand the numerous successes of the
QHD description of nuclear properties.
Nevertheless, finding an efficient, tractable, nonperturbative way to 
match the QCD Lagrangian to the long-range, strong-coupling, effective
field theory of QHD is still a major goal for the future.

\section*{Acknowledgments}

I am pleased to acknowledge fruitful collaborations with Sergei Ananyan,
Dick Furnstahl, Horst M\"uller, Hua-Bin Tang, and Dirk Walecka during 
the course of this work, which
was supported in part by the US Department of Energy under
Contract No.~DE--FG02--87ER40365.

\finalnewpage   


\end{document}